%% file: Diff_emu.tex
\pdfoutput=1
\newcommand*{\ATLASLATEXPATH}{latex/}
\documentclass[cernpreprint,texlive=2016,txfonts,UKenglish]{latex/atlasdoc} 
\usepackage{\ATLASLATEXPATH atlaspackage}
\usepackage{\ATLASLATEXPATH atlasbiblatex}
\usepackage{\ATLASLATEXPATH atlascontribute}
\usepackage{longtable}
\usepackage{\ATLASLATEXPATH atlasphysics}
\graphicspath{{logos/}{figures/}}
\usepackage{Diff_emu-defs}
\usepackage{pdflscape}

\input{Diff_emu-metadata}
\hypersetup{pdftitle={ATLAS document},pdfauthor={The ATLAS Collaboration}}

\begin{document}

\maketitle

\tableofcontents

\input{sections/introduction.tex}

\input{sections/detector.tex}

\input{sections/data_mc.tex}
\input{sections/object.tex}

\input{sections/reconstruction.tex}

\input{sections/unfolding.tex}

\input{sections/systematics.tex}

\input{sections/results.tex}

\FloatBarrier
\input{sections/conclusions.tex}

\section*{Acknowledgements}

\input{acknowledgements/Acknowledgements}

\clearpage

\printbibliography
\clearpage
\appendix

\newpage \input{atlas_authlist}

\end{document}

%% file: Diff_emu-metadata.tex
\AtlasTitle{Measurements of top-quark pair differential cross-sections in the $e\mu$
channel in \textit{pp} collisions at $\sqrt{s} = 13$~TeV using the ATLAS detector}

\author{The ATLAS Collaboration}

\AtlasRefCode{TOPQ-2016-04}

\PreprintIdNumber{CERN-PH-2016-220}

 \AtlasJournal{Eur.\ Phys.\ J.}
 \AtlasJournalRef{\EPJ C77 (2017) 299}
 \AtlasDOI{10.1140/epjc/s10052-017-4821-x}

\AtlasAbstract{%
  This article presents measurements of $t\bar{t}$ differential
  cross-sections in a fiducial phase-space region, using an integrated luminosity
  of 3.2 fb$^{-1}$ of proton--proton data at a centre-of-mass energy of $\sqrt{s} =
  13$~TeV recorded by the ATLAS experiment at the LHC in 2015.
  Differential cross-sections are measured as a function of the transverse momentum
  and absolute rapidity of the top quark, and of the transverse momentum, absolute rapidity
  and invariant mass of the $t\bar{t}$ system.
The $t\bar{t}$ events are selected by requiring one electron and one
  muon of opposite electric charge, and at least two jets, one of which must be tagged as
  containing a $b$-hadron. The measured differential cross-sections are compared
to predictions of next-to-leading order generators matched to parton showers and the measurements are
found to be consistent with all models within the experimental
  uncertainties with the exception of the \textsc{POWHEG}-Box + \textsc{HERWIG++} predictions, which
differ significantly from the data in both the transverse momentum of the top quark
and the mass of the $t\bar{t}$ system.

}

\AtlasJournal{EPJC}

%% file: sections/introduction.tex
\section{Introduction}
\label{sec:intro}

The top quark is the heaviest fundamental particle in the Standard
Model (SM) of particle physics. Understanding the production cross-section and kinematics of \ttbar pairs is
an important test of SM predictions. Furthermore, \ttbar production is often an
important background in searches for new physics and a detailed
understanding of this process is therefore crucial.

At the Large Hadron Collider (LHC), \ttbar\ pair production in proton--proton ($pp$)
collisions at a centre-of-mass energy of $\sqrt{s}=$~13 TeV occurs predominantly via gluon fusion
(90\%) with small contributions from \qqbar\ annihilation (10\%).
Significant progress has been made in the precision of the
calculations of the cross-section of this process,
both inclusive and differential. Currently,
calculations are available at next-to-next-to-leading order (NNLO) in
perturbative QCD, including the resummation of next-to-next-to-leading
logarithmic (NNLL) soft gluon terms~\cite{Cacciari:2011hy,Baernreuther:2012ws,Czakon:2012zr,Czakon:2012pz,Czakon:2013goa,Czakon:2011xx,Guzzi:2014wia,Kidonakis:2014pja,Ahrens:2010zv,Czakon:2015owf,Czakon:2016dgf}.

Differential cross-sections for \ttbar production have been
measured by the ATLAS~\cite{Aad:2014zka,Aad:2015mbv,Aad:2015eia} and CMS~\cite{cms_7, cms_8} experiments, in events containing either one or two charged leptons, at $\sqrt{s}=$~7 TeV and $\sqrt{s}=$~8 TeV. Measurements of \ttbar differential cross-sections at $\sqrt{s}=$~13 TeV have also been made at the CMS experiment~\cite{Khachatryan:2016mnb} in events containing one charged lepton. The integrated luminosity of 3.2 fb$^{-1}$ of $pp$ collision data collected by the ATLAS experiment at $\sqrt{s}=$~13~TeV allows the measurement of the differential
cross-section as a function of the kinematic variables of the
$t\bar{t}$ system in a different kinematic regime compared to the previous
LHC measurements. The inclusive cross-section has been measured
 at $\sqrt{s}=$~13 TeV by both the ATLAS~\cite{ATLAS_13TeV} and
CMS~\cite{CMS_13TeV, Khachatryan:2016kzg} experiments and was found to be in agreement with
the theoretical predictions. This article presents measurements of \ttbar differential cross-sections 
in terms of five different kinematic observables, both absolute and normalised to the fiducial 
cross-section. These observables are the transverse momentum of the top quark ($p_{\text T}(t)$),
the absolute rapidity of the top quark ($|y(t)|$), the transverse momentum of
the $t\bar{t}$ system ($p_{\text T}(t\bar{t})$), the absolute rapidity
of the $t\bar{t}$ system ($|y(t\bar{t})|$), and the invariant mass of
the $t\bar{t}$ system ($m(t\bar{t})$).  The distributions of these
variables are unfolded to the particle level in a fiducial volume. The $p_{\text T}(t)$ and $m(t\bar{t})$ 
observables are expected to be sensitive to the modelling of higher-order corrections
in QCD, whereas the rapidity of the top quark and $t\bar{t}$ system are
expected to have sensitivity to the parton distribution
functions (PDF) used in the simulations. The $p_{\text T}(t\bar{t})$ observable is sensitive
to the amount of gluon radiation in the event and can be useful for the
tuning of Monte Carlo (MC) generators. Top quarks and
anti-top quarks are measured in one combined distribution for
the $p_{\text T}(t)$ and $|y(t)|$ observables, rather than studying them separately.
The \ttbar\ system is reconstructed in events containing exactly one electron and one muon.
Events in which a $\tau$ lepton decays to an electron or muon are also included.

%% file: sections/detector.tex
\section{ATLAS detector}
\label{sec:detector}

The ATLAS detector~\cite{PERF-2007-01} at the LHC covers nearly the entire solid angle around the interaction point.
It consists of an inner tracking detector surrounded by a thin superconducting solenoid, electromagnetic and hadronic calorimeters,
and a muon spectrometer incorporating three large superconducting toroidal magnet systems.
The inner-detector system is immersed in a \SI{2}{\tesla} axial magnetic field 
and provides charged-particle tracking in the range $|\eta| < 2.5$.\footnote{ATLAS uses a right-handed coordinate system with its origin at the nominal interaction point (IP) in the centre of the detector and the $z$-axis along the beam pipe. The $x$-axis points from the IP to the centre of the LHC ring, and the $y$-axis points upwards. Cylindrical coordinates $(r, \phi)$ are used in the transverse plane, $\phi$ being the azimuthal angle around the $z$-axis. The pseudorapidity is defined in terms of the polar angle $\theta$ as $\eta = -\ln \tan(\theta / 2)$. Angular distance is measured in units of $\Delta R \equiv \sqrt{(\Delta\eta)^{2} + (\Delta\phi)^{2}}$.}

The high-granularity silicon pixel detector surrounds the collision region and provides 
four measurements per track. 
The closest layer, known as the Insertable B-Layer~\cite{Capeans:1291633},
was added in 2014 and provides high-resolution hits at small radius to improve
 the tracking performance.
The pixel detector is followed by the silicon microstrip tracker, which provides four three-dimensional measurement points per track.
These silicon detectors are complemented by the transition radiation tracker,
which enables radially extended track reconstruction up to $|\eta| = 2.0$. 
The transition radiation tracker also provides electron identification information 
based on the fraction of hits (typically 30 in total) 
passing a higher charge threshold indicative of transition radiation.

The calorimeter system covers the pseudorapidity range $|\eta| < 4.9$.
Within the region $|\eta|< 3.2$, electromagnetic calorimetry is provided by barrel and 
endcap high-granularity lead/liquid-argon (LAr) electromagnetic calorimeters,
with an additional thin LAr presampler covering $|\eta| < 1.8$
to correct for energy loss in material upstream of the calorimeters.
Hadronic calorimetry is provided by the steel/scintillator-tile calorimeter,
segmented into three barrel structures within $|\eta| < 1.7$, and two copper/LAr hadronic endcap calorimeters
 that cover $1.5 < |\eta| < 3.2$.
The solid angle coverage is completed with forward copper/LAr and tungsten/LAr calorimeter modules
optimised for electromagnetic and hadronic measurements respectively, in the region $3.1 < |\eta| < 4.9$.

The muon spectrometer comprises separate trigger and
high-precision tracking chambers measuring the deflection of muons in a magnetic field generated by superconducting air-core toroids.
The precision chamber system covers the region $|\eta| < 2.7$ with three layers of monitored drift tubes,
complemented by cathode strip chambers in the forward region, where the background is highest.
The muon trigger system covers the range $|\eta| < 2.4$ with resistive-plate chambers in the barrel, and thin-gap chambers in the endcap regions.

A two-level trigger system is used to select interesting events~\cite{PERF-2011-02, ATL-DAQ-PUB-2016-001}.
The Level-1 trigger is implemented in hardware and uses a subset of detector information
to reduce the event rate to a design value of at most \SI{100}{\kHz}.
This is followed by the software-based high-level trigger, which reduces the event rate to \SI{1}{\kHz}.

%% file: sections/data_mc.tex
\section{Data and simulation samples}
\label{sec:data_mc}

The $pp$ collision data used in this analysis were
collected during 2015 by ATLAS and correspond to an integrated
luminosity of 3.2 fb$^{-1}$ at $\sqrt{s} = 13$~TeV. 
The data considered in this analysis were collected under stable beam conditions, 
and requiring all subdetectors to be operational.
Each selected event includes additional interactions from, on average, 
14 inelastic {\textit pp} collisions in the same proton bunch crossing, as
well as residual detector signals from previous bunch crossings
with a 25 ns bunch spacing, collectively referred to as ``pile-up''. 
Events are required to pass a single-lepton trigger, either electron or
muon. Multiple triggers are used to select events: either triggers with low \pt
thresholds of 24 GeV that utilise isolation requirements to reduce the trigger rate, or
higher \pt thresholds of 50 GeV for muons or 60 and 120 GeV for electrons, with no isolation requirements to increase event
acceptance.

MC simulations are used to model background processes
and to correct the data for detector acceptance and
resolution effects. The ATLAS detector is simulated~\cite{fullsim} using
\GEANT4~\cite{Agostinelli:2002hh}. A ``fast
simulation''~\cite{ATLAS:1300517}, utilising parameterised showers in the calorimeter,
but with full simulation of the inner detector and muon spectrometer,
 is used in the samples generated to estimate \ttbar modelling uncertainties.
 Additional {\textit pp} interactions are generated using
\mbox{\PYTHIA\!8} (v8.186)~\cite{pythia8} and overlaid on signal and background
processes in order to simulate the effect of pile-up. The MC
simulations are reweighted to match the distribution of the average number of interactions
per bunch crossing that are observed in data. This process is referred to as ``pile-up
reweighting''. The same reconstruction algorithms and analysis procedures are applied to both data and MC simulation. 
Corrections derived from dedicated data samples are applied to the MC simulation in order to improve agreement with data. 

The nominal \ttbar\ sample is simulated using the
next-to-leading order (NLO) \POWHEG-Box (v2) matrix-element event generator \cite{powheg1,powheg2,powheg3}
using \mbox{\PYTHIA\!6}~(v6.427)~\cite{pythia6} for the parton shower (PS).
 \POWHEG-Box is interfaced to the CT10~\cite{CT10}
NLO PDF set while \mbox{\PYTHIA\!6} uses the CTEQ6L1 PDF set~\cite{CT6.1}. 
A set of tuned parameters called the Perugia 2012 tune~\cite{perugia2012}
 is used in the simulation of the underlying event. 
The ``$h_{\mathrm{damp}}$'' parameter, which controls the \pt\ of the first
additional gluon emission beyond the Born configuration, is set to the 
mass of the top quark ($m_{t}$). The main effect of this is to regulate the
high-\pt emission against which the \ttbar\ system recoils.
 The choice of this $h_{\mathrm{damp}}$ value was found to improve the
modelling of the \ttbar\ system kinematics with respect to data in previous
analyses~\cite{hdamp}. In order to investigate the effects of initial-
and final-state radiation, alternative \POWHEG-Box + \mbox{\PYTHIA\!6} samples are
generated with the renormalisation and factorisation scales varied by
a factor of 2 (0.5) and using low (high) radiation variations of the
Perugia 2012 tune and an $h_{\mathrm{damp}}$ value of $m_{t}$ ($2m_{t}$),
corresponding to less (more) parton-shower radiation~\cite{hdamp}, referred to as ``radHi'' and ``radLo''.
These variations were selected to cover the uncertainties in the measurements of 
differential distributions in $\sqrt{s}=7$~TeV data~\cite{Aad:2014zka}.
The $h_{\mathrm{damp}}$ value for the low radiation sample is not decreased
as it was found to disagree with previously published data.
Alternative samples are generated using \POWHEG-Box~(v2) and \MGMCatNLO~(v2.2.1)~\cite{amcatnlo}, 
referred to as \MGatNLO\ hereafter, both interfaced to \HERWIGpp~(v2.7.1)~\cite{herwig}, 
in order to estimate the effects of the choice of matrix-element event generator and parton-shower algorithm.
Additional \ttbar\ samples are generated for comparisons with unfolded data using
\SHERPA~(v2.2.0)~\cite{Sherpa}, \POWHEG-Box~(v2) + \mbox{\PYTHIA\!8} as well as
\POWHEG-Box~(v2) and \MGatNLO\ interfaced to \HERWIG\!7~\cite{herwig, Herwig7_2}.
In all \ttbar\ samples, the mass of the top quark is set to 172.5~GeV. These \ttbar\ 
samples are described in further detail in Ref.~\cite{hdamp}.

Background processes are simulated using a variety of MC
event generators. Single-top quark production in association with a $W$ boson ($Wt$) is simulated using \POWHEG-Box~v1 +
\mbox{\PYTHIA\!6} with the same parameters and PDF sets as those used for the nominal
\ttbar\ sample and is normalised to the theoretical cross-section~\cite{Kidonakis:2010ux}. 
The higher-order overlap with \ttbar\ production is addressed using the ``diagram removal'' (DR) generation
scheme~\cite{diagram_subtraction}. 
A sample generated using an alternative ``diagram subtraction'' (DS)
method is used to evaluate systematic uncertainties~\cite{diagram_subtraction}.

\SHERPA~(v2.1.1), interfaced to the CT10 PDF set, is used to
model Drell--Yan production, where the dominant contribution is from $Z/\gamma^*\rightarrow\tau^+\tau^-$.
For this process, \SHERPA\ calculates matrix elements at NLO for up to
two partons and at leading order (LO) for up to four partons using the OpenLoops~\cite{openloops}
and Comix~\cite{comix}
 matrix-element event generators. The matrix
elements are merged with the \SHERPA\ parton shower ~\cite{Schumann:2007mg} using the ME +
PS@NLO prescription~\cite{sherpanlo}. The total cross-section is
normalised to the NNLO predictions~\cite{ATL-PHYS-PUB-2016-003}.
\SHERPA~(v2.1.1) with the CT10 PDF set is also used to simulate
electroweak diboson production~\cite{ATL-PHYS-PUB-2016-002} ($WW$, $WZ$, $ZZ$), where both bosons
decay leptonically. For these samples, \SHERPA~calculates
matrix elements at NLO for zero additional partons, at LO for one to three
additional partons (with the exception of $ZZ$ production, for which
the one additional parton is also at NLO), and using PS for all parton
multiplicities of four or more. All samples are normalised 
using the cross-section computed by the event generator.

Events with \ttbar\ production in association with a vector boson are
simulated using \MGatNLO\ + \mbox{\PYTHIA\!8}~\cite{madgraph}, using the
NNPDF2.3 PDF set and the A14 tune, as described in Ref.~\cite{ATL-PHYS-PUB-2016-005}. 

Background contributions containing one prompt lepton and one misidentified (``fake'')
lepton, arising from either a heavy-flavour hadron decay, photon conversion,
jet misidentification or light-meson decay, are estimated using samples from
MC simulation. The history of the stable particles in the generator-level record is
used to identify fake leptons from these processes by identifying leptons that originated from hadrons. The majority ({\raise.17ex\hbox{$\scriptstyle\mathtt{\sim}$}}90\%) of
fake-lepton events originate from the single-lepton \ttbar\ process, with
smaller contributions arising from $W$~+~jets and \ttbar\ + vector-boson
events. $W$~+~jets events are simulated using \POWHEG-Box~+~\mbox{\PYTHIA\!8} with
 the CT10 PDF set and the AZNLO tune~\cite{AZNLO:2014}. 
 The $t$-channel single-top quark process is generated using \POWHEG-Box~v1 + \mbox{\PYTHIA\!6} 
 with the same parameters and PDF sets as those used for the nominal \ttbar sample.
\textsc{EvtGen} (v1.2.0)~\cite{EvtGen} is used for the heavy-flavour hadron decays in all samples.
 Other possible processes with fake leptons, such as multi-jet and Drell--Yan production, are negligible for the event selection
used in this analysis.

%% file: sections/object.tex
\section{Object and event selection}
\label{sec:object}

This analysis utilises reconstructed electrons, muons, jets and missing transverse momentum (with magnitude \met). Electron candidates are identified 
by matching an inner-detector track to an isolated energy deposit 
in the electromagnetic calorimeter, within the fiducial region of transverse momentum $\pt > 25$~GeV
and pseudorapidity $|\eta| < 2.47$. Electron candidates are excluded if the calorimeter cluster is within the transition region between the barrel and the endcap of the
electromagnetic calorimeter, $1.37 < |\eta| < 1.52$. Electrons are selected using a multivariate algorithm and 
are required to satisfy a likelihood-based quality criterion, in order to provide high
efficiency and good rejection of fake electrons~\cite{Aad:2014fxa,ATLAS-CONF-2016-024}.
Electron candidates must have tracks
that pass the requirements of transverse impact parameter significance\footnote{The transverse impact parameter significance
is defined as $d_0^\text{sig} = d_0 / \sigma_{d_0}$, where $\sigma_{d_0}$ is the uncertainty in the transverse impact parameter $d_0$.}
 $|d_0^\text{sig}|<5$ and
longitudinal impact parameter $|z_0 \sin\theta| < 0.5$~mm. 
Electrons must pass isolation requirements based on inner-detector tracks and 
topological clusters in the calorimeter which depend on $\eta$ and $p_{\textrm T}$.
These requirements result in an isolation efficiency of 95\% for an electron \pt of 25~GeV and
99\% for an electron \pt above 60~GeV when determined in simulated $Z\rightarrow e^{+}e^{-}$ events. The fake-electron rate 
determined in simulated \ttbar events is 2\%. 
Electrons that share a track with a muon are discarded.
Double counting of electron energy deposits as jets is
prevented by removing the closest jet within $\Delta R = 0.2$ of a reconstructed electron.
Following this, the electron is discarded if a jet exists within $\Delta R = 0.4$ of the electron 
to ensure sufficient separation from nearby jet activity.

Muon candidates are identified from muon-spectrometer tracks that match tracks in the inner
detector, with $\pt > 25$~GeV and $|\eta| < 2.5$~\cite{Aad:2016jkr}. The
tracks of muon candidates are required to have a transverse impact parameter significance
$|d_0^\text{sig}|<3$  and longitudinal impact parameter $|z_0 \sin\theta| < 0.5$~mm.
Muons must satisfy quality criteria and isolation requirements based on inner-detector tracks and topological 
 clusters in the calorimeter which depend on $\eta$ and $p_{\textrm T}$. 
These requirements reduce the contributions from fake muons and 
provide the same efficiency as for electrons when determined in simulated \ttbar events. 
Muons may leave energy deposits in the calorimeter that could be misidentified as a jet, so
jets with fewer than three associated tracks are removed if they are within $\Delta R = 0.4$ 
of a muon.  Muons are discarded if they are separated from the nearest jet by $\Delta R < 0.4$ to
reduce the background from muons from heavy-flavour hadron decays inside jets.

Jets are reconstructed with the anti-$k_t$ algorithm~\cite{Cacciari:2005hq,Cacciari:2008gp}, using a radius parameter of $R = 0.4$,
from topological clusters of energy deposits in the calorimeters. Jets are accepted within the range $\pt > 25$~GeV 
and $|\eta| < 2.5$, and are calibrated using simulation with corrections derived from data~\cite{ATL-PHYS-PUB-2015-015}.
Jets likely to originate from pile-up are suppressed using a multivariate jet-vertex-tagger (JVT)~\cite{ATLAS-CONF-2014-018,Aad:2015ina}
for candidates with $\pt < 60$~GeV and $|\eta| < 2.4$. 
Jets are identified as candidates for containing $b$-hadrons 
 using a multivariate discriminant~\cite{ATL-PHYS-PUB-2015-022}, which uses track impact parameters,
track invariant mass, track multiplicity and secondary vertex information to discriminate $b$-jets from light-quark or gluon jets (light jets). 
The average $b$-tagging efficiency is 76\%, with a purity of 90\%, for $b$-jets in simulated dileptonic \ttbar events.

\met is reconstructed using calibrated electrons, muons and jets~\cite{ATL-PHYS-PUB-2015-027}, 
where the electrons and muons are required to satisfy the selection criteria above.
Tracks associated with the primary vertex are used for the computation of \met
from energy not associated with electrons, muons or jets. The primary
vertex is defined as the vertex with the highest sum of $p_{\textrm T}^2$ of tracks associated with it.

Signal events are selected by requiring exactly one electron and one muon of opposite
electric charge, and at least two jets, at least one of which must be $b$-tagged. No requirements are made on 
the \met in the event. Using this selection, 85\% of events
are expected to be \ttbar\ events. The other
processes that pass the signal selection are Drell--Yan ($Z/\gamma^{*}\rightarrow\tau^{+}\tau^{-}$), 
diboson and single-top quark ($Wt$) production and fake-lepton events.

\begin{figure}
\centering
\begin{tabular}{cc}
\subfloat[\label{fig:emu_electron_pt}]{\includegraphics[width=0.45\textwidth]{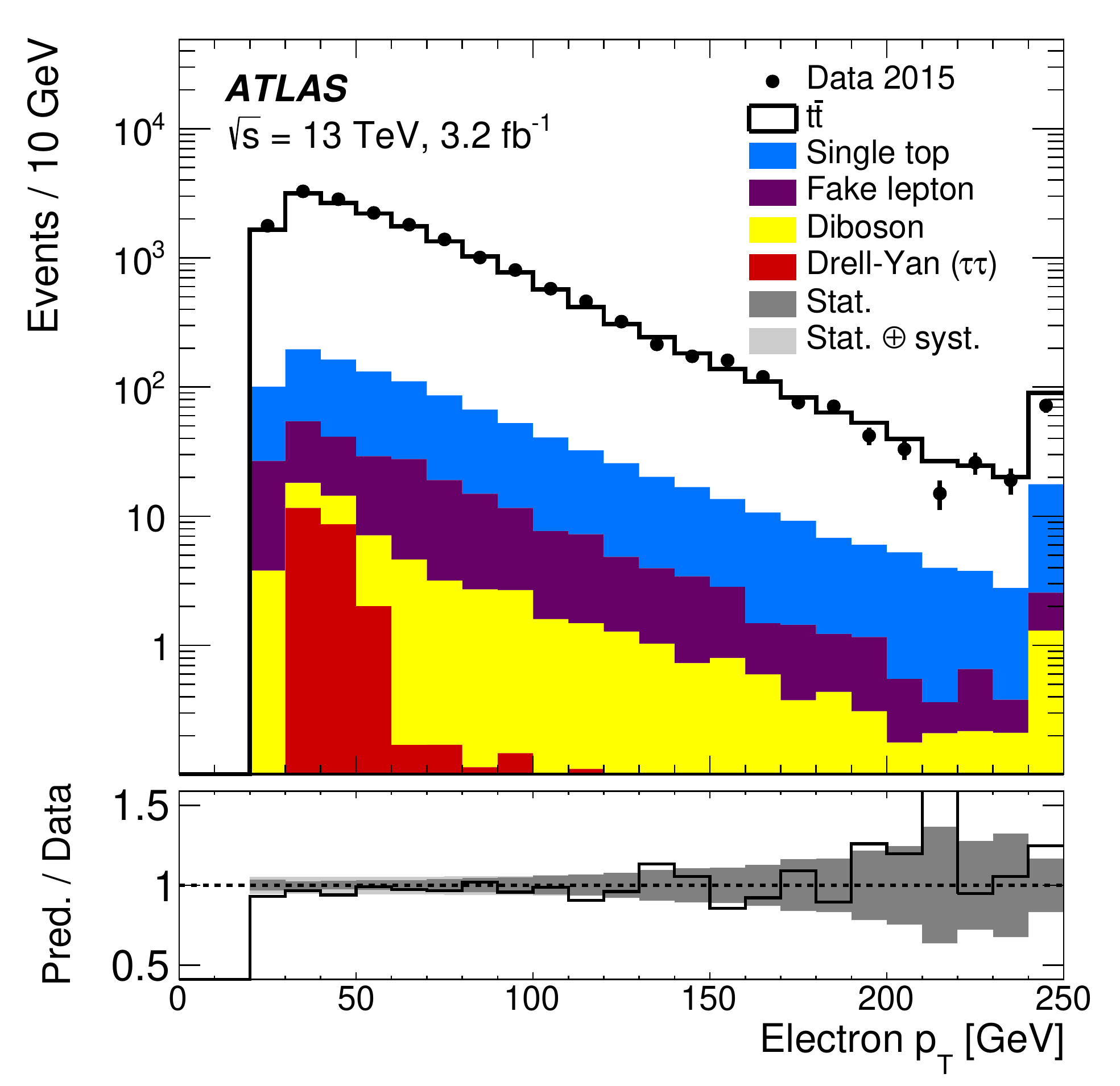}}&
\subfloat[\label{fig:emu_muon_pt}]{\includegraphics[width=0.45\textwidth]{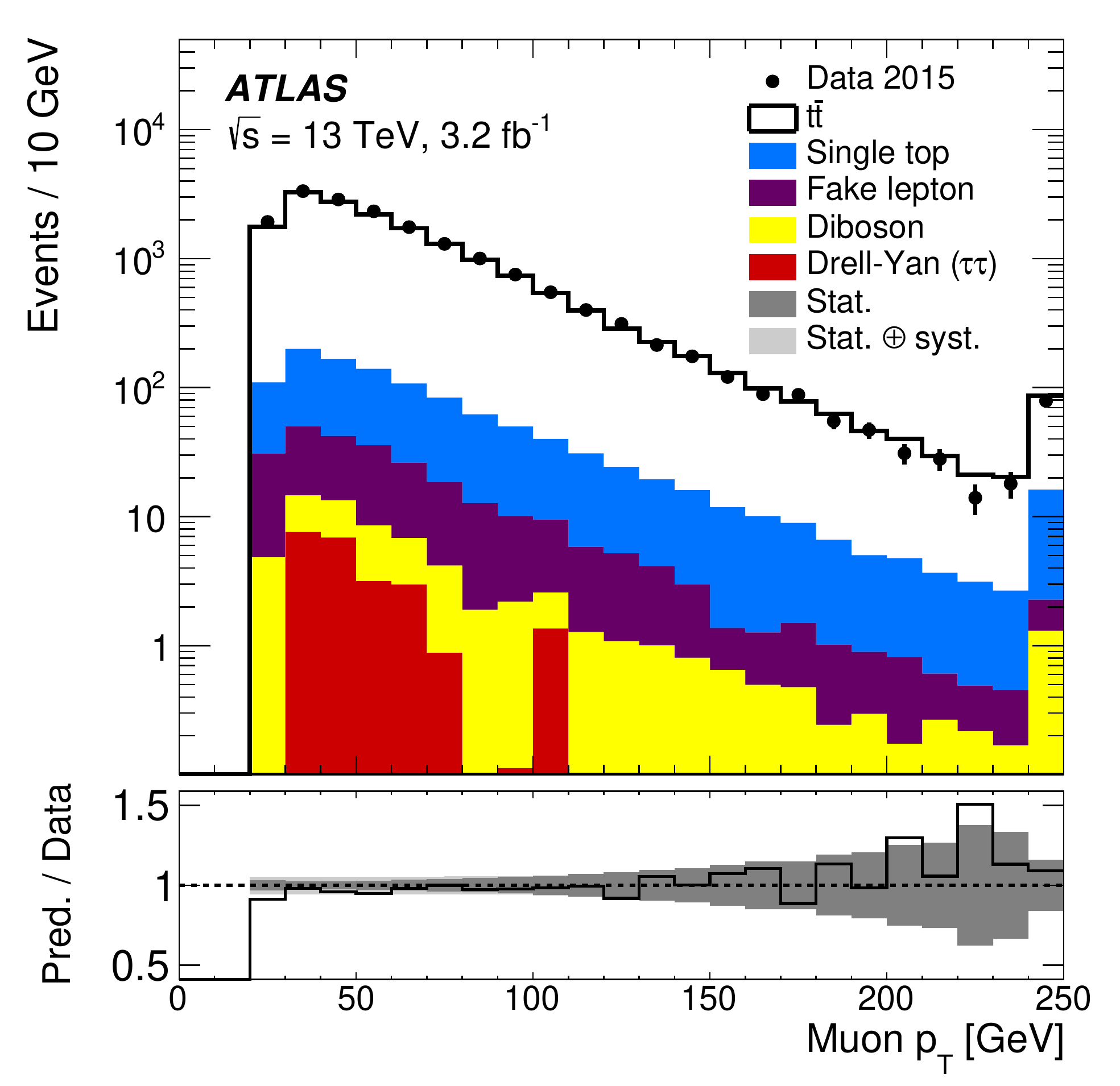}} \\
\subfloat[\label{fig:emu_bjet_pt}]{\includegraphics[width=0.45\textwidth]{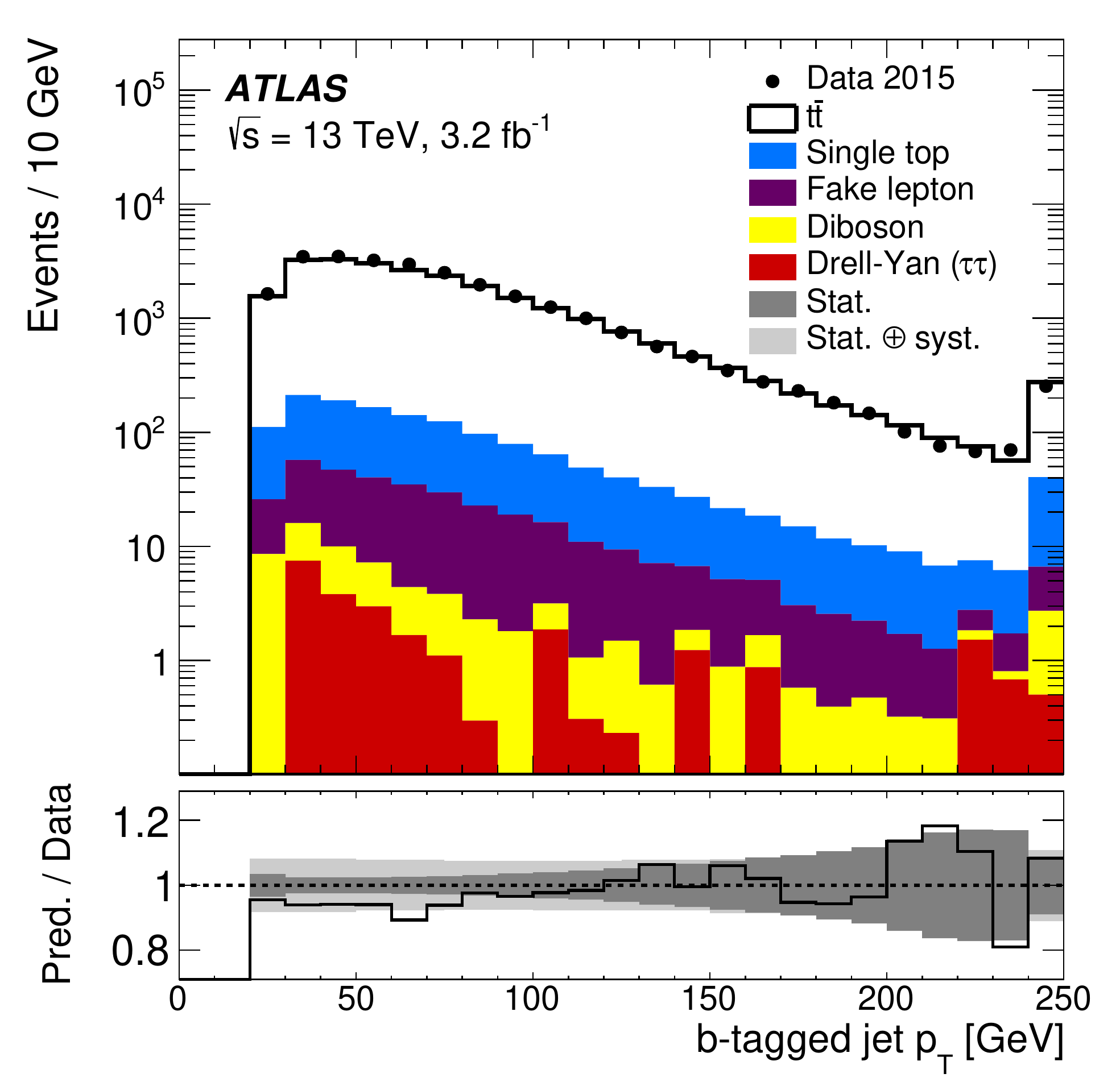}} &
\subfloat[\label{fig:emu_met}]{\includegraphics[width=0.45\textwidth]{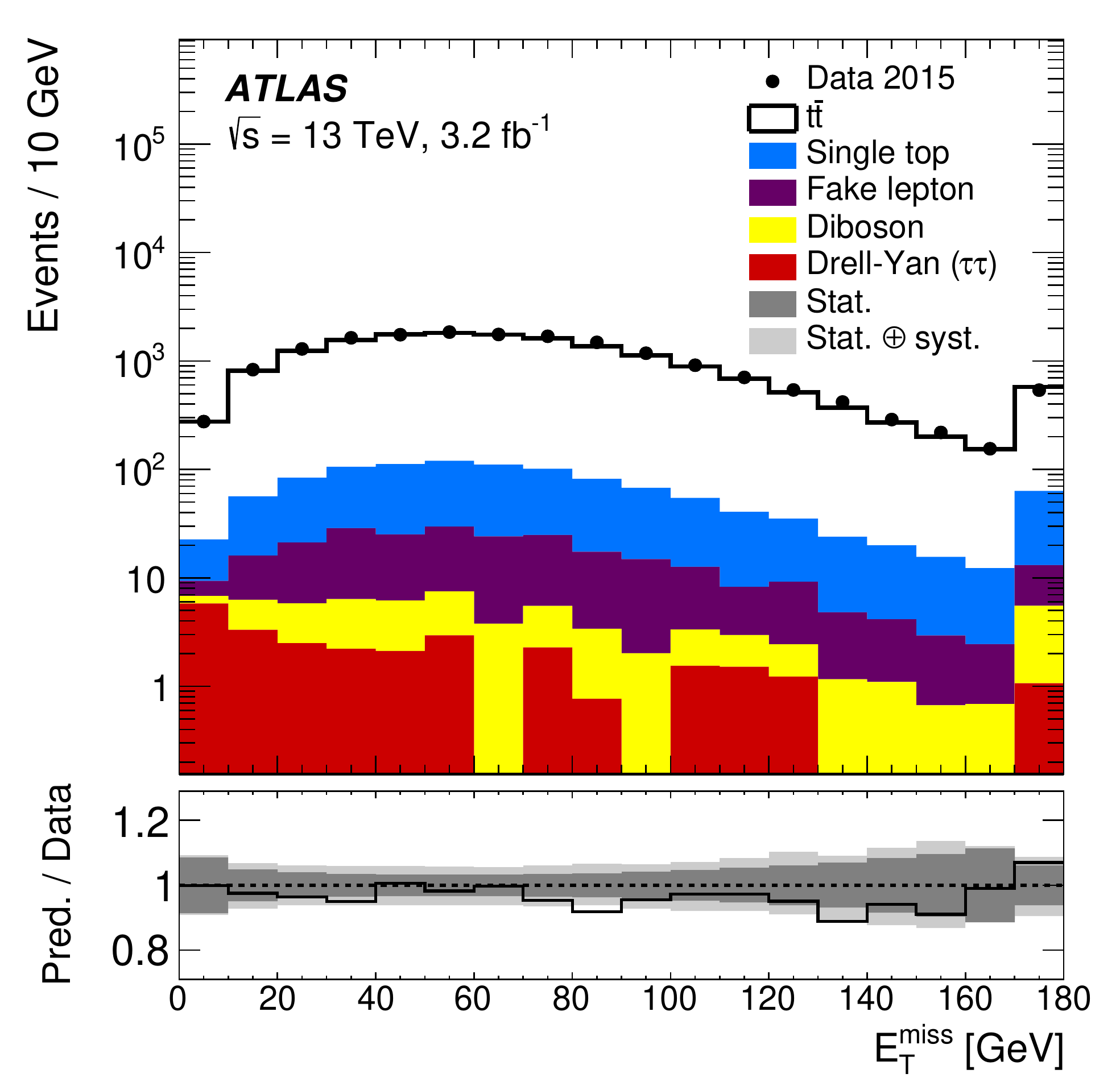}}\\
\end{tabular}
\caption{Kinematic distributions for the electron \pt (a), muon \pt (b),
  $b$-jet \pt (c), and \met (d) for the $e^{\pm}\mu^{\mp}$ signal selection. 
  In all figures, the rightmost bin also contains events that are above the $x$-axis range.
  The dark uncertainty bands in the ratio plots represent the statistical uncertainties while
  the light uncertainty bands represent the statistical, systematic and luminosity
  uncertainties added in quadrature.
 The uncertainties quoted include
  uncertainties from leptons, jets, missing transverse momentum,
   background modelling and pile-up modelling. They do not
  include uncertainties from PDF or signal \ttbar\ modelling.}
\label{fig:emu_selection}
\end{figure}

The event yields after the signal selection are listed in Table~\ref{tab:emu_selection_nuw}.
The number of events observed in the signal region exceeds the prediction, but the
excess is within the uncertainties. Distributions of
lepton and jet \pt and \met are shown in
Figure~\ref{fig:emu_selection}. The $t\bar{t}$ contribution is
normalised using the predicted cross-section, calculated with the
Top++2.0 program at next-to-next-to-leading order in perturbative QCD,
including soft-gluon resummation to next-to-next-to-leading-logarithm order~\cite{Czakon:2011xx}
and assuming a top-quark mass of 172.5 GeV. 
The data and prediction agree within the total uncertainty for all
distributions. The $p_{\textrm T}$ observables show a small deficit in the
simulation prediction at low $p_{\textrm T}$ which was found to be
correlated with the modelling of the top-quark $p_{\textrm T}$.

\begin{table}[h]
\centering
\begin{tabular}{ l |
 S[table-format=5.1,
   table-figures-uncertainty=1]|
 S[table-format=5.1,
   table-figures-uncertainty=1]
}
\toprule
Process &  {Signal region} & {Signal region + NW} \\
\midrule

$Z/\gamma^{*}\rightarrow\tau^{+}\tau^{-}$         & 22    \pm 9   & 10    \pm 8 \\
   Diboson                                      & 44    \pm 4   & 17    \pm 2 \\
Fake lepton                                     & 200   \pm 60  & 150   \pm 50 \\
      $Wt$                                      & 860   \pm 60  & 480   \pm 40 \\
$t\bar{t}$                                      & 15800 \pm 900 & 13300 \pm 800 \\
\midrule
  Expected                                      & 17000 \pm 900 & 13900 \pm 800 \\
  Observed                                      & 17501         & 14387 \\

\bottomrule
\end{tabular}
\caption{Event yields in the signal selection, and after requiring that
neutrino weighting (NW) reconstructs the event.
  The quoted uncertainties include
  uncertainties from leptons, jets, missing transverse momentum,
  luminosity, statistics, background modelling and pile-up modelling. They do not
  include uncertainties from PDF or signal \ttbar\ modelling. The results and uncertainties are rounded according
  to recommendations from the Particle Data Group (PDG).}
\label{tab:emu_selection_nuw}
\end{table}
 
Particle-level objects are constructed using generator-level information in the MC simulation,
using a procedure intended to correspond as closely as possible to the reconstructed object and event 
selection. Only objects in the MC simulation with a
lifetime longer than $3 \times 10^{-11}$~s (stable) in the generator-level information are used. Particle-level 
electrons and muons are identified as those originating from a
$W$-boson decay, including those via intermediate $\tau$ leptons. The four-momenta of each
electron or muon is summed with the four-momenta of all radiated photons, excluding those from hadron decays, within a cone of size $\Delta R = 0.1$,
and the resulting objects are required to have $\pt > 25$~GeV and $|\eta| < 2.5$. 
Particle-level
jets are constructed using stable particles, with the exception of
selected particle-level electrons and muons and particle-level
neutrinos originating from $W$-boson decays,
using the anti-$k_t$ algorithm with a radius parameter of $R = 0.4$, in the region 
$\pt > 25$~GeV and $|\eta| < 2.5$. 
Intermediate $b$-hadrons in the MC decay chain history are clustered in the stable-particle jets
with their energies set to zero.
 If, after clustering, a particle-level jet contains one or more of these ``ghost'' $b$-hadrons, the
 jet is said to have originated from a $b$-quark.
This technique is referred to as ``ghost matching''~\cite{ghost}.
Particle-level \met is calculated using the vector transverse-momentum sum of all neutrinos in the event,
 excluding those originating from hadron decays, either directly or via a $\tau$ lepton.

Events are selected at the particle level in a fiducial phase space region with similar requirements to  the phase space region at reconstruction level. Events are selected by requiring exactly one particle-level electron and one particle-level muon of opposite
electric charge, and at least two particle-level jets, at least one of which must originate from a $b$-quark.

%% file: sections/reconstruction.tex
\section{Reconstruction}
\label{sec:reconstruction}

The $t$, $\bar{t}$, and \ttbar are reconstructed using both the
particle-level objects and the reconstructed objects in order to measure their
kinematic distributions. The reconstructed system is built using the neutrino
weighting (NW) method~\cite{NeutrinoWeighting}. 

Whereas the individual four-momenta of the two neutrinos in the final state
are not directly measured in the detector, the sum of
their transverse momenta is measured as \met. The absence of the
measured four-momenta of the two neutrinos leads to an under-constrained
system that cannot be solved analytically. However, if additional
constraints are placed on the mass of the top-quark, the mass of the $W$
boson, and on the pseudorapidities of the two neutrinos, the system can be
solved using the following equations:
\begin{equation}
 \label{eq:nus}
 \begin{aligned}
(\ell_{1,2} + \nu_{1,2})^2 &= m^2_W = (80.2~\text{\GeV})^2\text{,} \\
(\ell_{1,2} + \nu_{1,2} + b_{1,2})^2 &= m_t^2 = (172.5~\text{\GeV})^2\text{,}\\
\eta(\nu),~\eta(\bar{\nu}) &= \eta_{1},~\eta_{2} \text{,}\\
\end{aligned}
\end{equation}
where $\ell_{1,2}$ are the charged leptons, $\nu_{1,2}$ are the neutrinos, and $b_{1,2}$ are the 
$b$-jets (or jets), representing four-momentum vectors, and $\eta_{1},~\eta_{2}$
are the assumed $\eta$ values of the two neutrinos. 
Since the neutrino $\eta$'s are unknown, many different assumptions of their
values are tested. The possible values for $\eta(\nu)$
and $\eta(\bar{\nu})$ are scanned between $-5$ and 5 in steps of 0.2.

With the assumptions about $m_t$, $m_W$, and values for $\eta(\nu)$ and $\eta(\bar{\nu})$, Equation (\ref{eq:nus}) can now be solved, leading to two possible solutions for each
assumption of $\eta(\nu)$ and $\eta(\bar{\nu})$. Only real solutions
without an imaginary component are considered. The observed \met value in each event is used to determine which solutions are more 
likely to be correct.
A ``reconstructed'' \met value resulting from the neutrinos for each solution
is compared to the \met observed in the event.
If this reconstructed \met value matches the observed \met value in the event, then
the solution with those values for $\eta(\nu)$ and $\eta(\bar{\nu})$
is likely to be the correct one. A weight is introduced in order to
quantify this agreement:
\begin{equation}
 \label{eq:weight}
w = \exp\Bigg(\frac{-\Delta E^2_x}{2\sigma_x^2}\Bigg) \cdot \exp \Bigg( \frac{-\Delta E^2_y}{2\sigma^2_y} \Bigg) \text{,}
\end{equation}
where $\Delta E_{x,y}$ is the difference between the 
missing transverse momentum computed from Equation (\ref{eq:nus}) and the observed missing transverse momentum
in the $x$--$y$ plane and $\sigma_{x,y}$ is the resolution of the
observed \met\ in the detector in the $x$--$y$ plane. 
The assumption for $\eta(\nu)$ and $\eta(\bar{\nu})$ that gives the highest weight
is used to reconstruct the $t$
and $\bar{t}$ for that event. The \met resolution is taken to be 15 GeV for both the $x$ and $y$
directions~\cite{ATL-PHYS-PUB-2015-027}. This choice has little effect on
which solution is picked in each event.
The highest-weight solution remains the same
regardless of the choice of $\sigma_{x,y}$.

In each event, there may be more than two jets and therefore many
possible combinations of jets to use in the kinematic reconstruction. In
addition, there is an
ambiguity in assigning a jet to the $t$ or to the $\bar{t}$ candidate. In events with only one $b$-tagged jet, the $b$-tagged jet and the highest-$p_{\textrm T}$ non-$b$-tagged jet are used to reconstruct the $t$ and $\bar{t}$, whereas in events with two or more $b$-tagged jets, the two $b$-tagged jets with the highest weight from the $b$-tagging algorithm are used.

Equation (\ref{eq:nus}) cannot always be solved for a particular
assumption of $\eta(\nu)$ and $\eta(\bar{\nu})$. This can be caused by
misassignment of the input objects or
through mismeasurement of the input object four-momenta. It is also possible that the assumed $m_t$ 
is sufficiently different from the true value to prevent a valid solution for that event. To mitigate these
effects, the assumed value of $m_t$ is varied between the
values of 168 and 178 GeV, in steps of 1 GeV, and the \pt of the measured jets 
are smeared using a Gaussian function with a width of 10\% of
their measured \pt. This smearing is repeated 20 times.
 This allows the NW algorithm to
shift the four-momenta (of the electron, muon and the two jets)
 and $m_t$ assumption to see if a solution can be found. The solution
 which produces the highest $w$ is taken as the reconstructed system.

\begin{figure}
\centering
\subfloat[]{\label{fig:emu_top_pt}\includegraphics[width=0.39\textwidth]{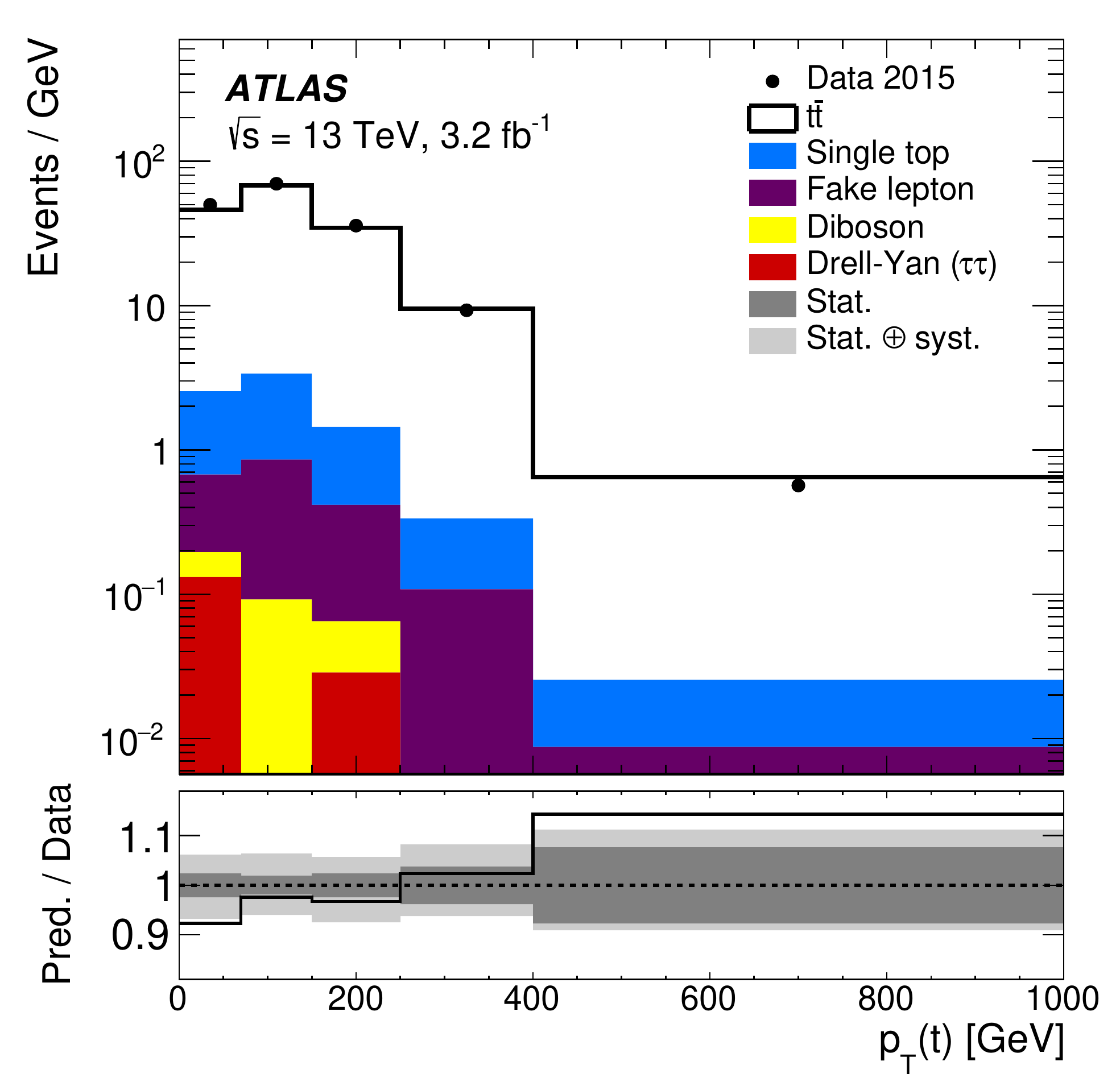}}
\subfloat[]{\label{fig:emu_top_y}\includegraphics[width=0.39\textwidth]{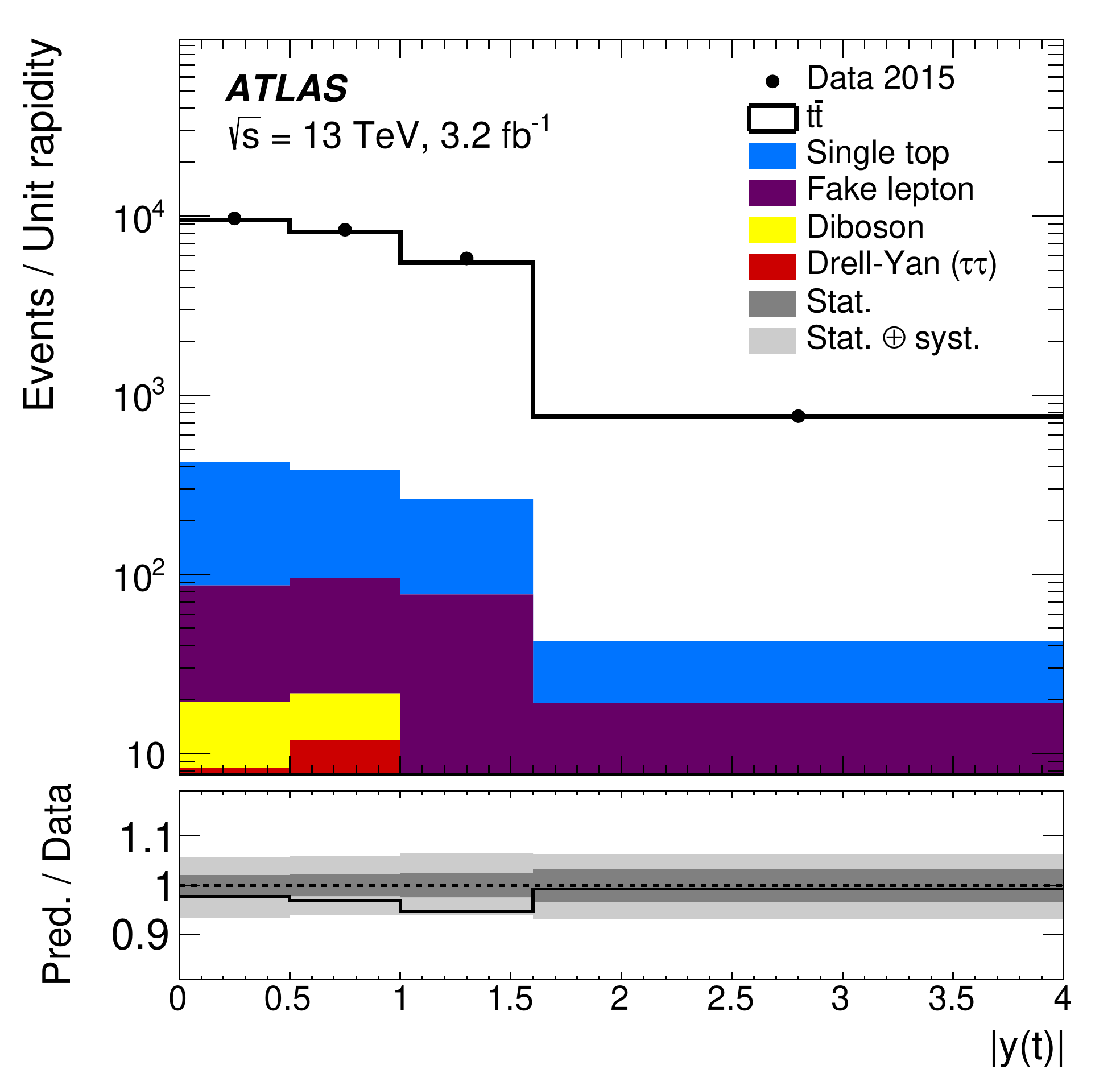}}\\
\subfloat[]{\label{fig:emu_ttbar_pt}\includegraphics[width=0.39\textwidth]{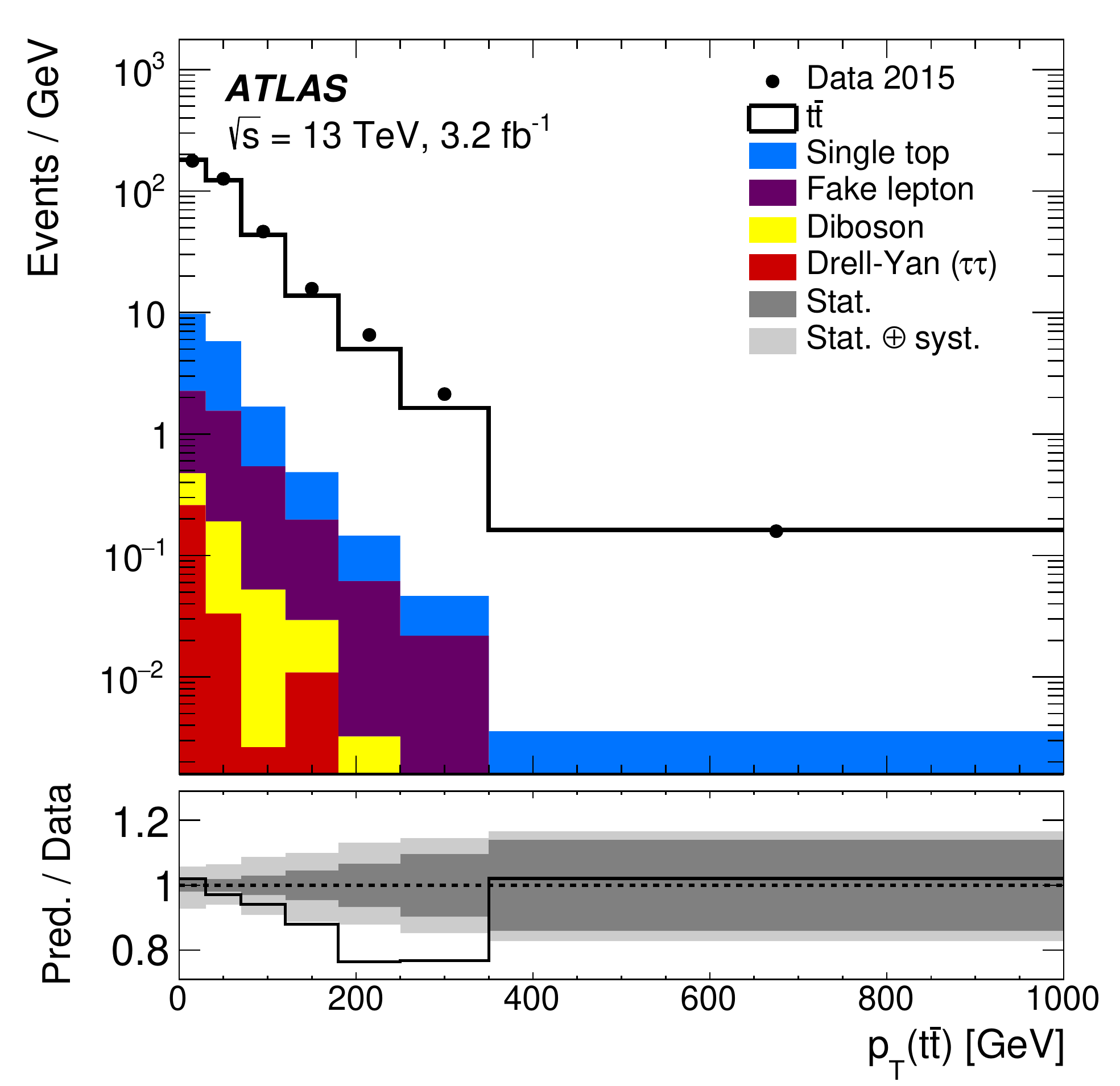}}
\subfloat[]{\label{fig:emu_ttbar_y}\includegraphics[width=0.39\textwidth]{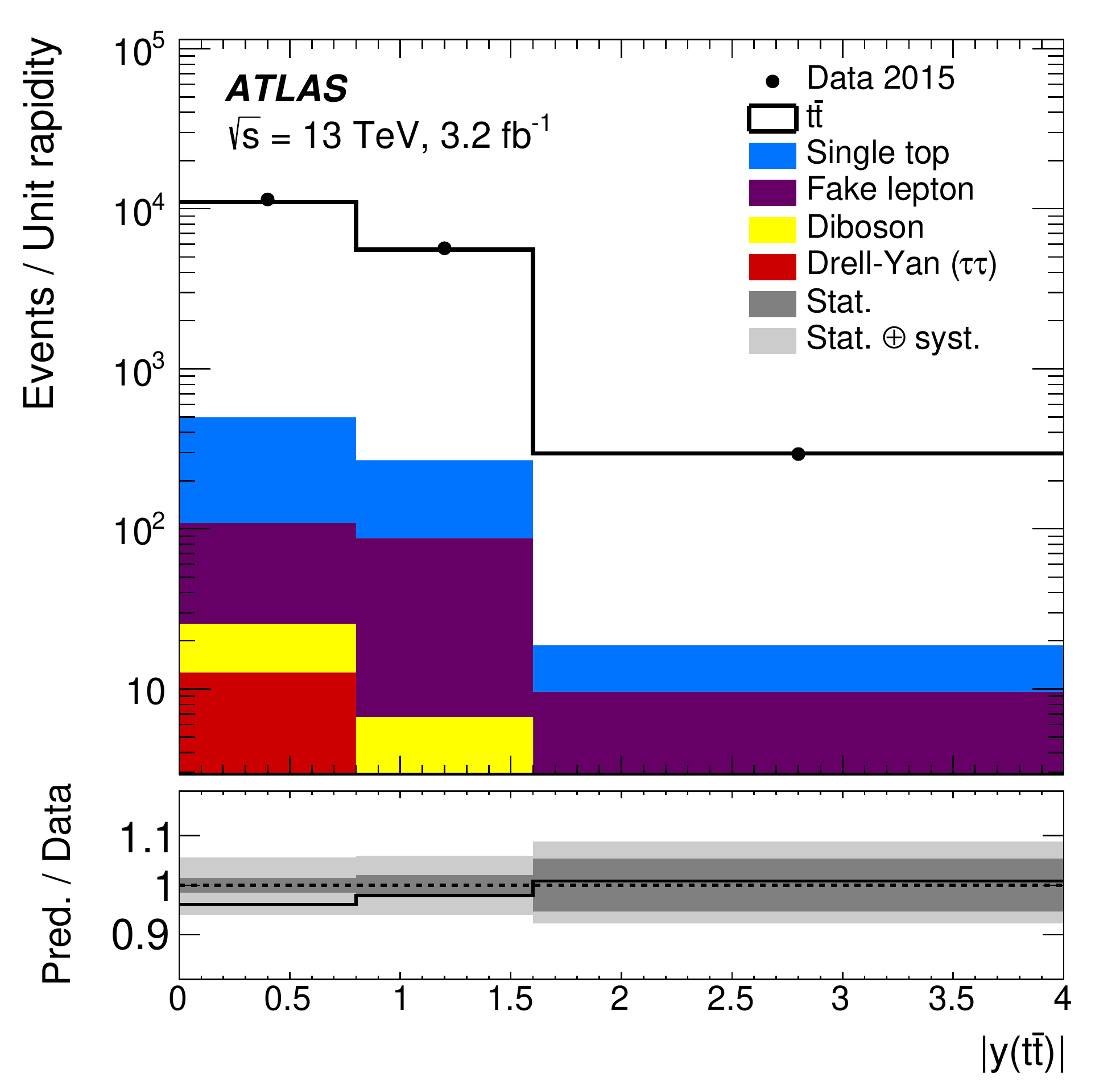}}\\
\subfloat[]{\label{fig:emu_ttbar_m}\includegraphics[width=0.39\textwidth]{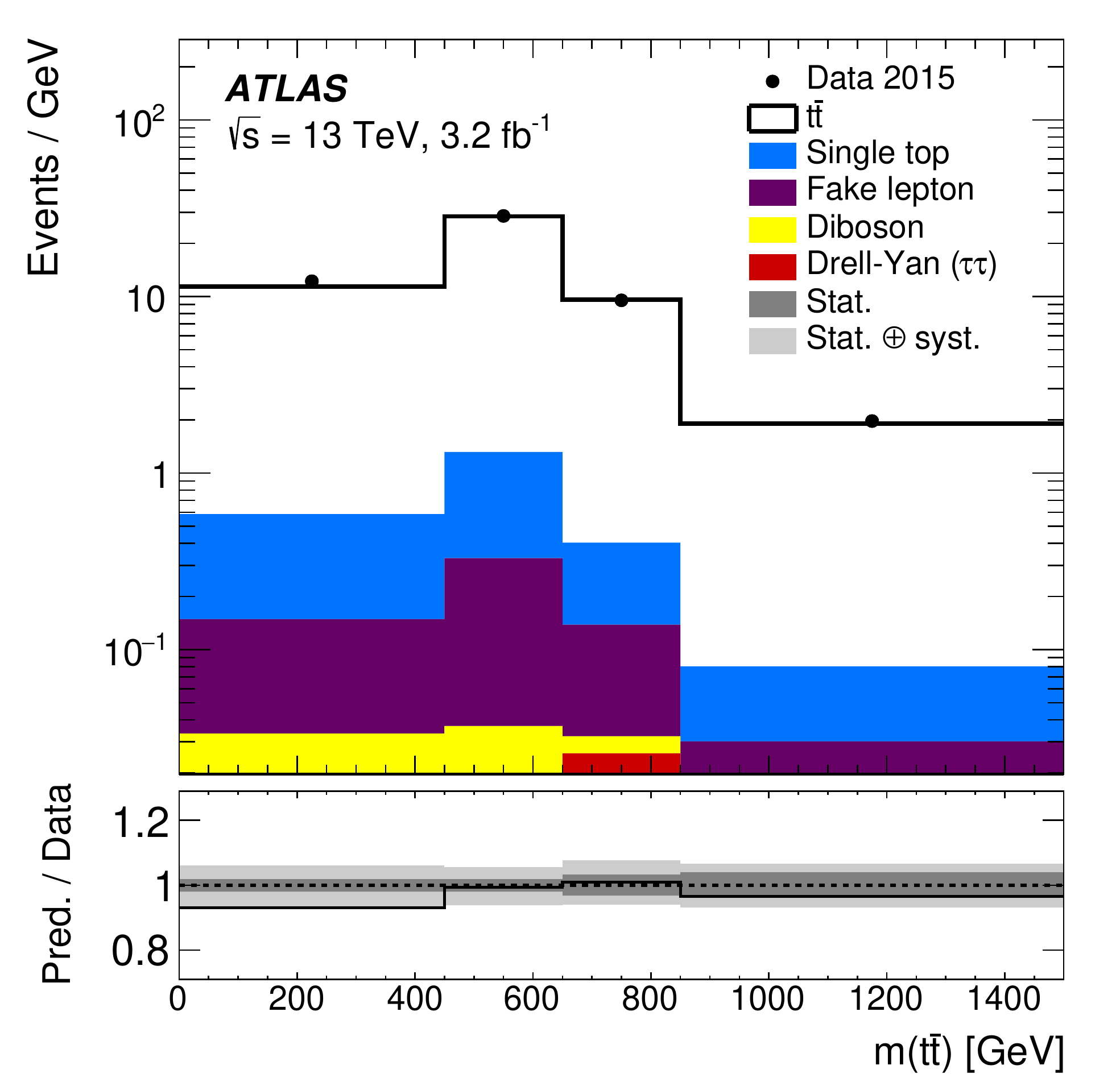}}
\caption{Kinematic distributions for the $\pt(t)$ (a), $|y(t)|$ (b),
  $\pt(\ttbar)$ (c), \ttbar $|y_{\ttbar}|$ (d), and $m(t\bar{t})$ (e) after reconstruction of the \ttbar system. In all figures, the
  rightmost bin also contains events that are above the $x$-axis
  range. The uncertainty bands represent the statistical uncertainties
  (dark) and the statistical, systematic
  and luminosity uncertainties added in quadrature (light). The uncertainties
  quoted include uncertainties on leptons, jets, \met,
  background and pile-up modelling, and luminosity. They do not
  include uncertainties on PDF or signal \ttbar\ modelling.}
\label{fig:emu_selection_nuW_top} 
\end{figure}

For a fraction of events, even smearing does not help to find a solution. Such
events are not included in the signal selection and are counted as an inefficiency of the reconstruction. 
For the signal \ttbar\ MC
samples, the inefficiency is {\raise.17ex\hbox{$\scriptstyle\mathtt{\sim}$}}20\%. Due to
the implicit assumptions about the $m_t$ and $m_W$, the reconstruction
inefficiency found in simulated background samples is much higher ({\raise.17ex\hbox{$\scriptstyle\mathtt{\sim}$}}40\% for
$Wt$ and Drell--Yan processes) and leads to a suppression of background events.
Table~\ref{tab:emu_selection_nuw} shows the event yields before and after reconstruction in the signal region.  
The purity of \ttbar events increases after
reconstruction. The distributions of the experimental observables after reconstruction
 are shown in Figure~\ref{fig:emu_selection_nuW_top}.

Particle-level~$t$,~$\bar{t}$,~and $t\bar{t}$ objects are reconstructed
following the prescriptions from the LHCTopWG, with the exception that only events with at least one $b$-tagged
jet are allowed. Events are required to have exactly two leptons of opposite-sign electric charge 
(one electron and one muon), and at least two jets.
The $t$ and $\bar{t}$ are reconstructed by
considering the two particle-level neutrinos with the highest \pt and
the two particle-level charged leptons. The charged leptons and
the neutrinos are paired such that $|m_{\nu_1,\ell_1} - m_W| +
|m_{\nu_2,\ell_2} - m_W|$ is minimised. These pairs are then used as
pseudo $W$ bosons and are paired with particle-level jets such that $|m_{W_1,j_1} - m_t| +
|m_{W_2,j_2} - m_t|$ is minimised, where at least one of the jets must
be $b$-tagged. In cases where only one particle-level $b$-jet is present,
the particle-level jet with the highest \pt among the non-$b$-tagged jets
 is used as the second jet.
 In cases with two particle-level $b$-jets, both are
taken. In the rare case of events with more than two particle-level $b$-jets,
 the two highest-\pt particle-level $b$-jets are
used. The particle-level \ttbar\ object is constructed using the sum
of the four-momenta of the particle-level $t$ and $\bar{t}$.

%% file: sections/unfolding.tex
\section{Unfolding}
\label{sec:unfolding}

To obtain the absolute and normalised differential cross-sections 
in the fiducial phase space region (see Section~\ref{sec:object}) with respect to
the \ttbar system variables, the distributions are unfolded to
particle level using an iterative Bayesian method~\cite{bayesian} implemented 
in the \textsc{RooUnfold} package~\cite{roounfold}.
In the unfolding, background-subtracted data are 
corrected for detector acceptance and resolution effects as well as for the efficiency to 
pass the event selection requirements in order to obtain the absolute differential cross-sections. 
The fiducial differential cross-sections are divided by the measured total cross-section, 
obtained by integrating over all bins in the differential distribution, in order to 
obtain the normalised differential cross-sections.

The differential cross-sections are calculated using the equation:
\begin{equation}
\frac{\text{d}\sigma_{\ttbar}}{\text{d}X_i} = \frac{1}{\mathcal{L} \cdot \mathcal{B} \cdot \Delta X_i \cdot \epsilon_i} \cdot \sum_{j} R^{-1}_{ij} \cdot \epsilon^{\text{fid}}_j \cdot (N_j^{\text{obs}}-N_j^{\text{bkg}}) \text{,}
\end{equation}
where $i$ indicates the bin for the observable $X$, $\Delta X_i$ is the width of bin $i$, 
$\mathcal{L}$ is the integrated luminosity, $\mathcal{B}$ is the branching ratio of 
the process ($\ttbar \rightarrow b \bar{b} e^{\pm} \nu_e \mu^{\mp} \nu_{\mu}$), $R$ is the response matrix, 
$N_j^\text{obs}$ is the number of observed events in data in bin $j$, and $N_j^\text{bkg}$ is the estimated number of 
background events in bin $j$. The efficiency parameter, $\epsilon_i$ ($\epsilon^{\textrm fid}_j$), is used to correct for events passing the reconstructed (fiducial) event selection but not the fiducial (reconstructed) selection.

\begin{figure}
\centering
\subfloat[]{\label{fig:a}\includegraphics[width=0.45\textwidth]{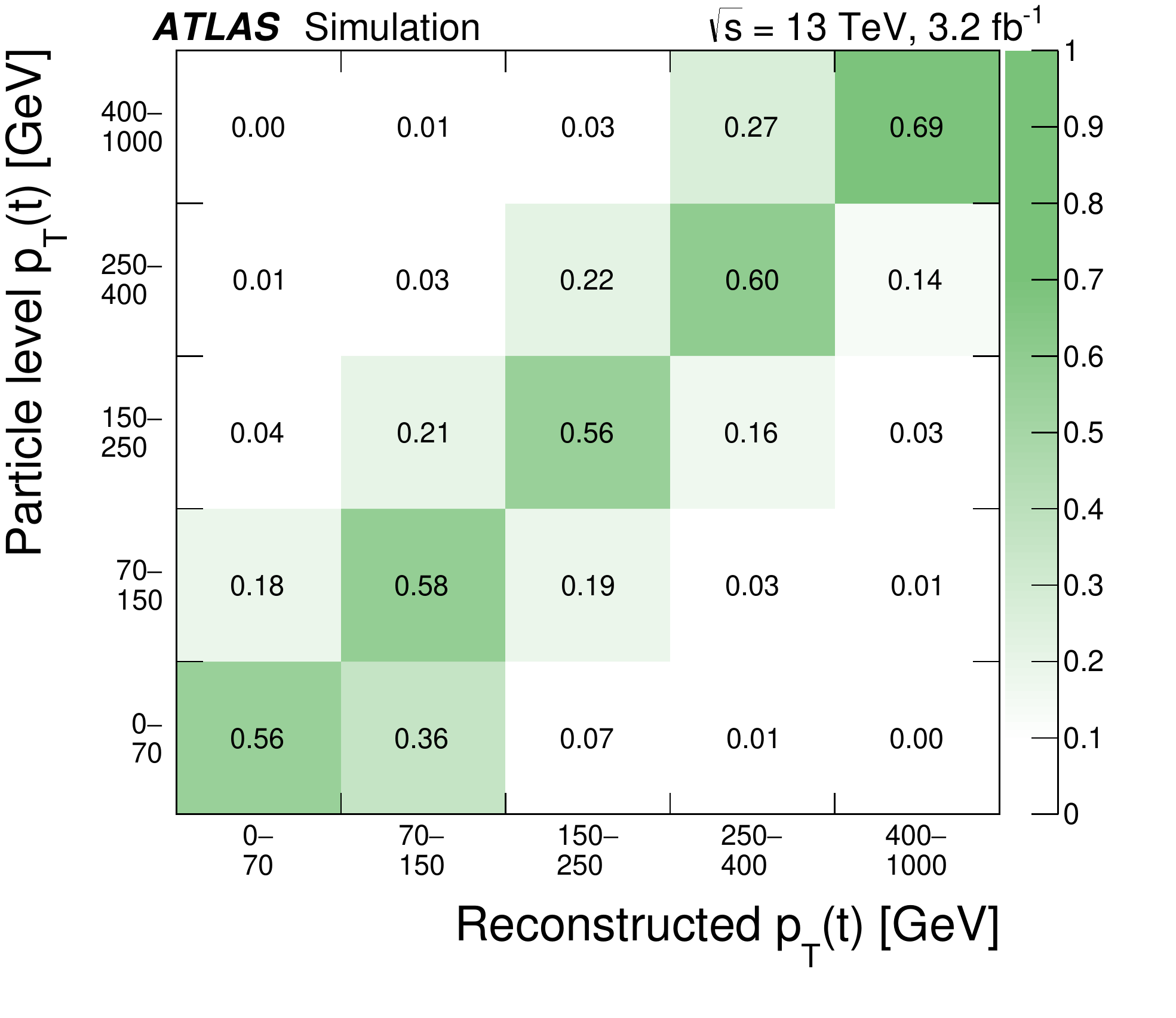}}
\subfloat[]{\label{fig:b}\includegraphics[width=0.45\textwidth]{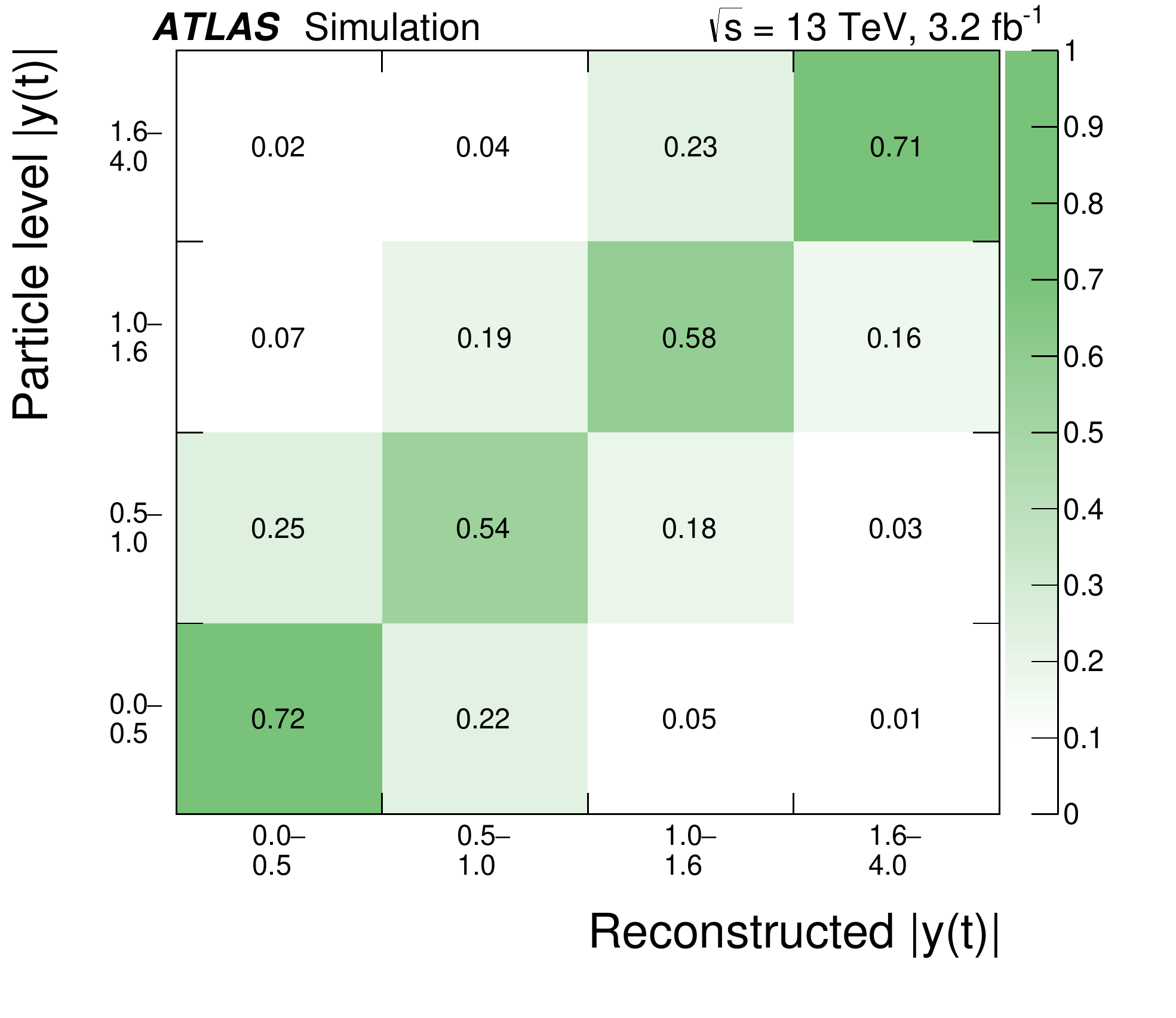}}\\
\subfloat[]{\label{fig:c}\includegraphics[width=0.45\textwidth]{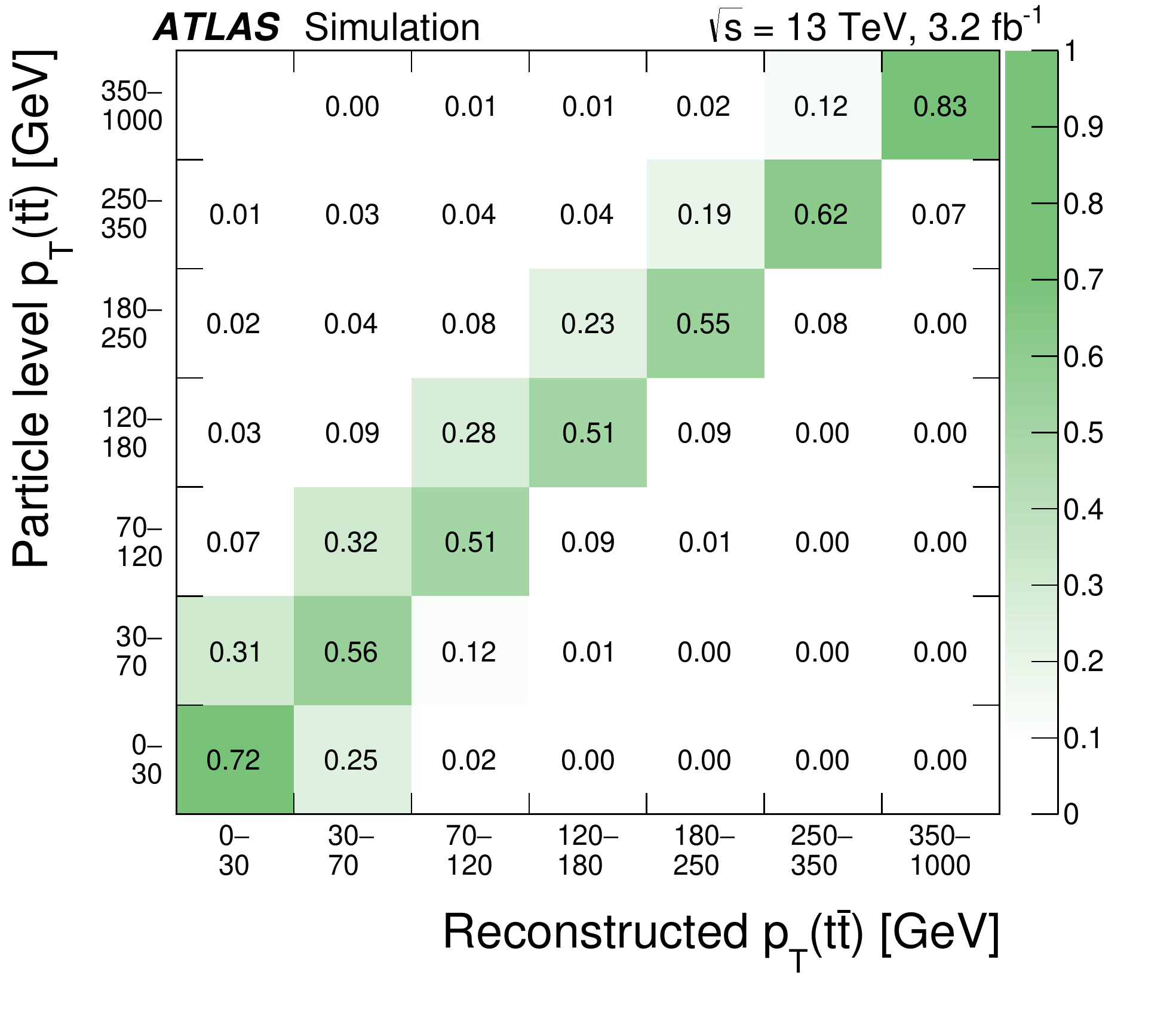}}
\subfloat[]{\label{fig:d}\includegraphics[width=0.45\textwidth]{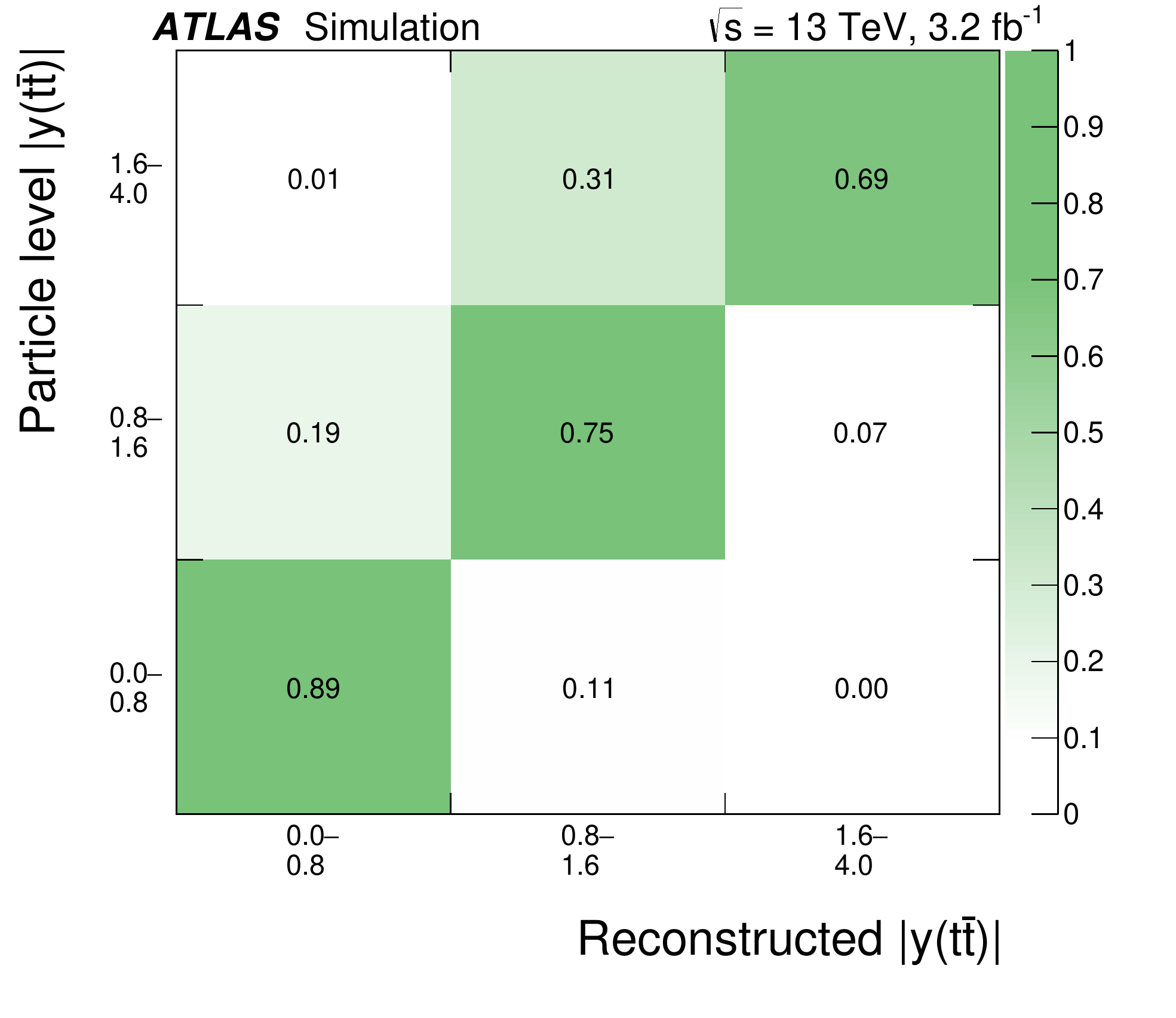}}\\
\subfloat[]{\label{fig:e}\includegraphics[width=0.45\textwidth]{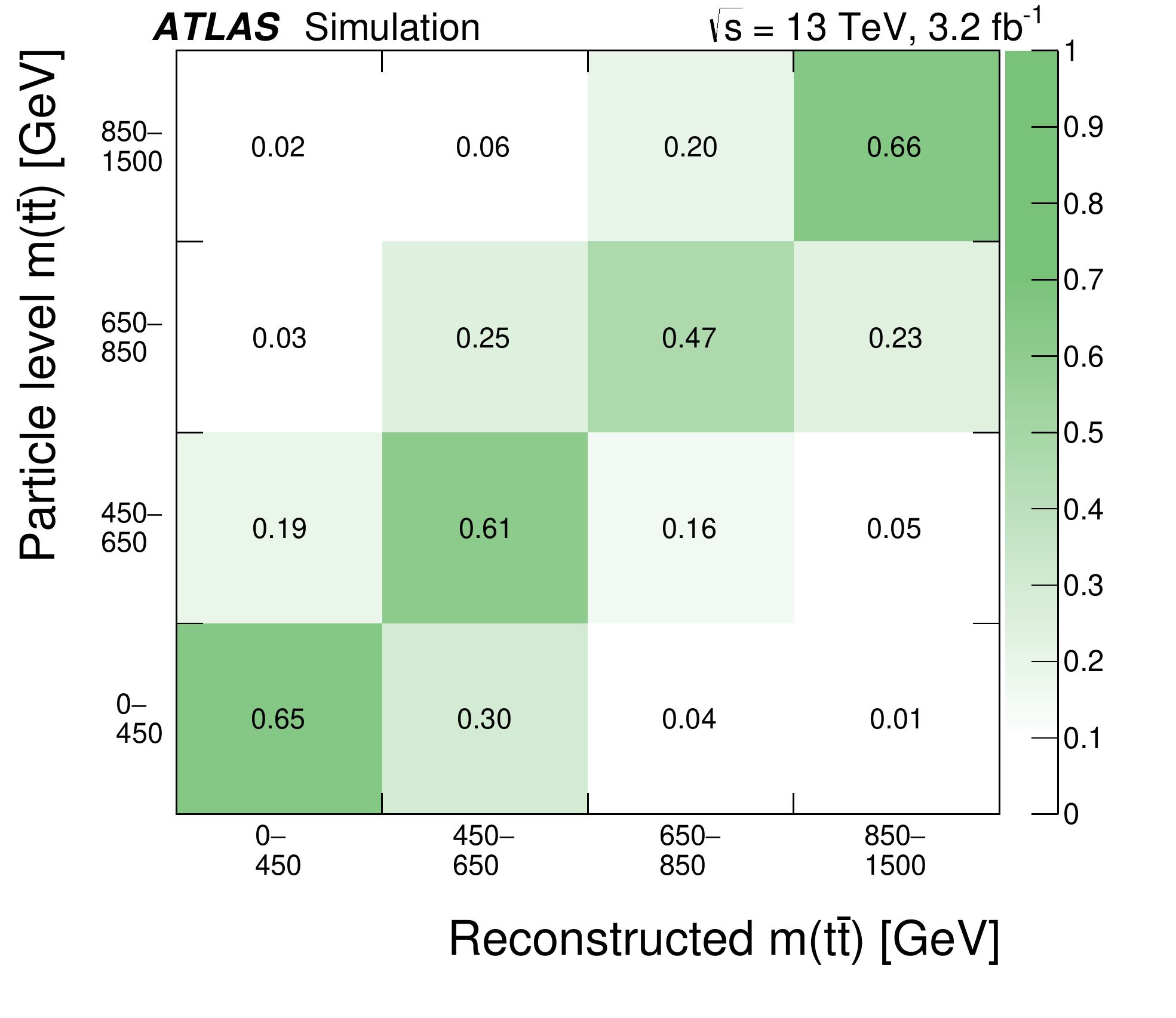}}
\caption{The response matrices for the observables obtained from the nominal \ttbar MC, normalised by row to unity.
Each bin shows the probability for a particle-level event in bin $j$ to be observed in a reconstruction-level bin $i$. White corresponds to 0 probability and the darkest green to a probability of one, where the other probabilities lie in between those shades.
}
\label{fig:migration_em_particle}
\end{figure}

The response matrix, $R$, describes the detector response, and is determined by 
mapping the bin-to-bin migration of events from particle level to reconstruction level
 in the nominal \ttbar MC simulation.
Figure~\ref{fig:migration_em_particle} 
shows the response matrices that are used for each experimental observable, normalised such that the sum of entries
in each row is equal to one. 
The values represent the fraction of events at particle level in bin $i$ that are reconstructed in bin $j$ at reconstruction level. 

The binning for the observables is chosen such that approximately half of the events are reconstructed in 
the same bin at reconstruction level as at the particle level (corresponding to a value of approximately 0.5 in 
the diagonal elements of the migration matrix). 
Pseudo-data are constructed by randomly sampling events from the nominal \ttbar MC sample,
 to provide a number of events similar to the number expected from data. 
 These pseudo-data are used to establish the stability of unfolding with respect to the choice of binning with pull tests.
 The binning choice must result in pulls consistent with a mean of zero and a standard deviation of one, within uncertainties. The choice of binning does not introduce any bias or underestimation of the statistical uncertainties.
 The number of iterations used in the iterative Bayesian unfolding is also optimised using pseudo-experiments. Iterations are 
performed until the $\chi^2$ per degree of freedom, calculated by comparing the unfolded pseudo-data to the corresponding generator-level 
distribution for that pseudo-data set, is less than unity. The optimum number of iterations is determined to be six.
 Tests are performed to establish that the unfolding procedure is able to successfully unfold distributions other than
those predicted by the nominal MC simulation.

%% file: sections/systematics.tex
\section{Systematic uncertainties}
\label{sec:systematics}

\begin{figure}
\centering
\begin{tabular}{cc}
\subfloat[\label{fig:syst_unc_top_pt}]{\includegraphics[width=0.495\textwidth]{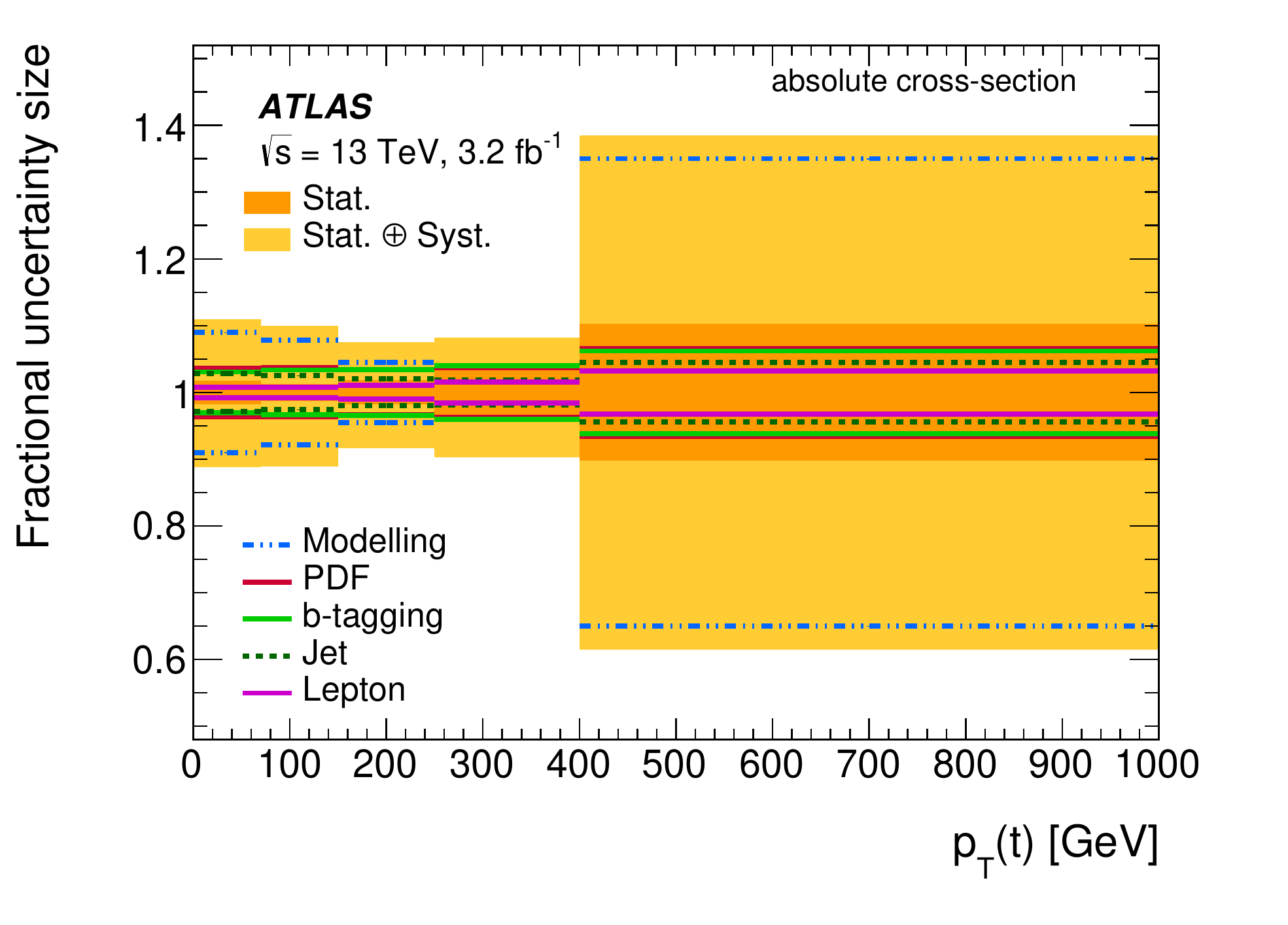}}&
\subfloat[\label{fig:syst_unc_top_pt_norm}]{\includegraphics[width=0.495\textwidth]{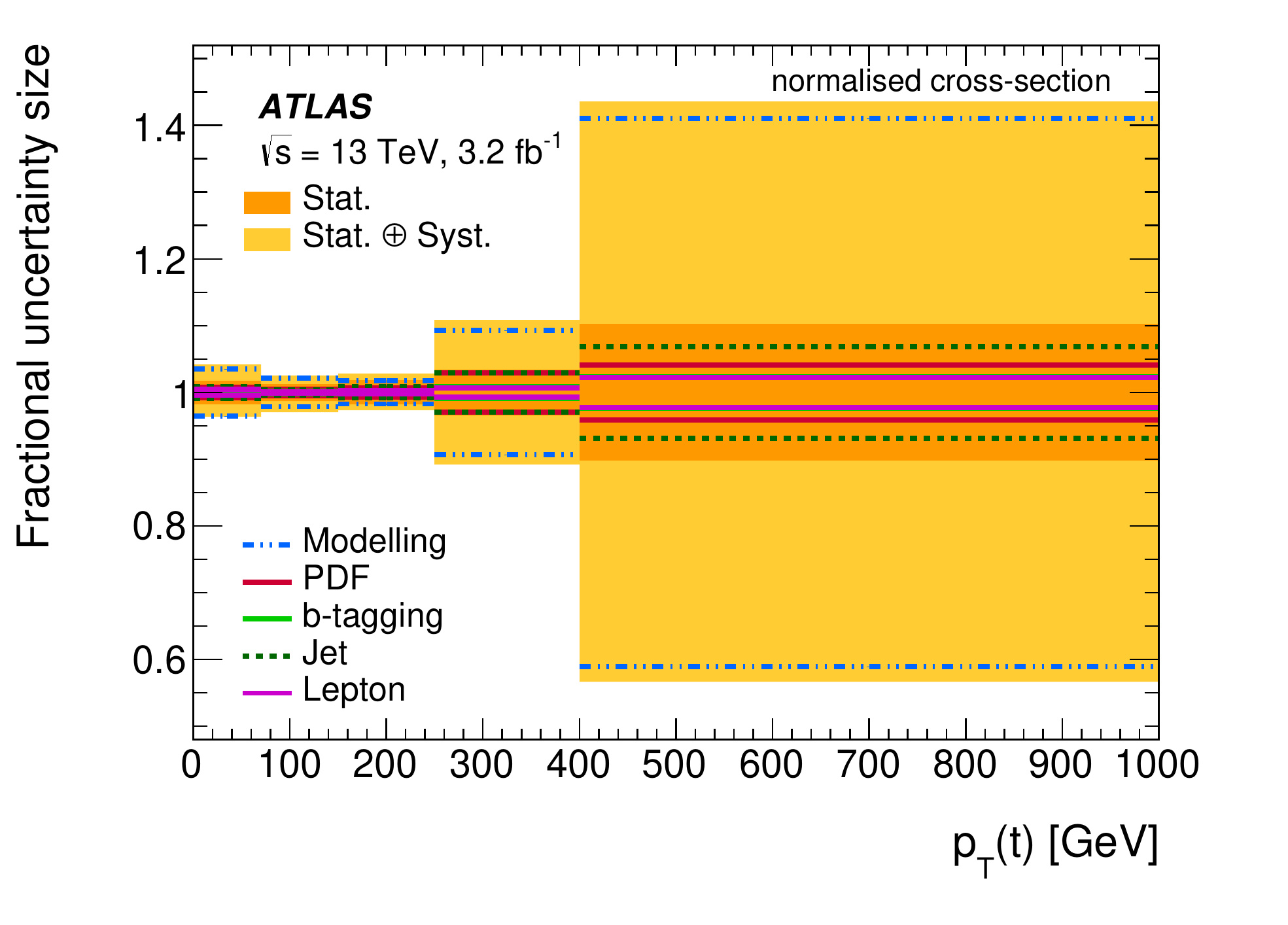}}\\
\end{tabular}
\caption{Summary of the fractional size of the absolute (a) and normalised (b) fiducial differential cross-sections as a function of $p_{\text{T}}(t)$. Systematic uncertainties which are symmetric are represented by solid lines and asymmetric uncertainties are represented by dashed or dot-dashed lines. Systematic uncertainties from common sources, such as modelling of the $t\bar{t}$ production, have been grouped together. Uncertainties due to luminosity or background modelling are not included. The statistical and total uncertainty sizes are indicated by the shaded bands.}
\label{fig:systematics}
\end{figure}

The measured differential cross-sections are affected by
systematic uncertainties arising from detector response, signal modelling, and background
modelling. The contributions from various sources of uncertainty are
described in this section.
Summaries of the sources of uncertainty for the absolute and normalised differential
cross-sections for the $p_{\textrm T}(t)$ are presented in Tables~\ref{tab:syst_small_top_pt} and
\ref{tab:syst_small_top_pt_normalised}. The total systematic uncertainties are calculated by 
summing all of the individual systematic uncertainties in quadrature and the total uncertainty is
calculated by summing the systematic and statistical uncertainties in quadrature.
The effect of different groups of systematic uncertainties is shown graphically for $p_{\textrm T}(t)$ in Figure~\ref{fig:systematics}.

\subsection{Signal modelling uncertainties}

The following systematic uncertainties related to the modelling of the \ttbar\
system in the MC generators are considered: the choice of 
matrix-element generator, the hadronisation model, the
choice of PDF, and the amount of initial- and final-state radiation.

Each source is estimated by using a different MC sample in the unfolding
procedure. In particular, a chosen baseline MC sample is unfolded 
using response matrices and corrections
derived from an alternative sample. The difference between the
unfolded distribution in the baseline sample and the true distribution in the baseline sample
is taken as the systematic uncertainty due to the signal modelling.

The choice of NLO generator (MC generator) affects the kinematic properties
 of the simulated \ttbar events and the reconstruction
efficiencies. To estimate this uncertainty, a comparison between \POWHEG-Box and \MGatNLO~(both using \HERWIGpp for
the parton-shower simulation) is performed, with the \POWHEG-Box sample used as the
baseline. The resulting systematic shift is used to define a
symmetric uncertainty, where deviations from the nominal sample are
also considered to be mirrored in the opposite direction, resulting in
equal and opposite symmetric uncertainties (called symmetrising).

To evaluate the uncertainty arising from the choice of parton-shower algorithm,
a sample generated using \POWHEG-Box~+~\mbox{\PYTHIA6} is compared to the alternative sample 
 generated with \mbox{\POWHEG-Box}~+~\HERWIGpp, where both samples use ``fast simulation''.
 The resulting uncertainty is symmetrised.
 The choices of NLO generator and parton-shower algorithm are dominant sources of
 systematic uncertainty in all observables.

The uncertainty due to the choice of PDF is evaluated using the
 PDF4LHC15 prescription~\cite{pdf4lhc}. The prescription utilises 100
eigenvector shifts derived from fits to the CT14~\cite{CT14}, MMHT~\cite{CT14} and NNPDF3.0~\cite{NNPDF}
PDF sets (PDF4LHC 100). The nominal MC sample used in the analysis is generated
using the CT10 PDF set. Therefore, the uncertainty is taken to
be the standard deviation of all eigenvector variations summed in quadrature with the
difference between the central values of the CT14
and CT10 PDF sets (PDF extrapolation). The resulting uncertainty is
symmetrised. Both PDF-based uncertainties contribute as one of the
dominant systematic uncertainties.

Uncertainties arising from varying the amount of 
initial- and final-state radiation (radiation scale), which alters the
jet multiplicity in events and the transverse momentum of the \ttbar\ system, are estimated
by comparing the nominal \POWHEG-Box~+~\mbox{\PYTHIA6} sample to 
samples generated with high and low radiation settings, as discussed in Section~\ref{sec:data_mc}.
 The uncertainty is taken as the difference between the nominal and the
increased radiation sample, and the nominal and the decreased
radiation sample. The initial- and final-state radiation is a
significant source of uncertainty in the absolute cross-section
measurements but only a moderate source of uncertainty in the
normalised cross-sections.

\subsection{Background modelling uncertainties}
The uncertainties in the background processes are assessed by repeating
the full analysis using pseudo-data sets and by varying the background predictions by one
standard deviation of their nominal values. The difference between the
nominal pseudo-data set result and the shifted result is taken as the
systematic uncertainty. 

Each background prediction has an uncertainty associated with its
theoretical cross-section. The cross-section for the $Wt$ process is varied by $\pm$5.3\%~\cite{Kidonakis:2010ux},
 the diboson cross-section is varied by $\pm$6\%, and the Drell--Yan $Z/\gamma^*\rightarrow\tau^+\tau^-$
 background is varied by $\pm$5\% based on studies of different MC
 generators. A 30\% uncertainty is assigned to the
normalisation of the fake-lepton background based on comparisons
 between data and MC simulation in a fake-dominated control region, which
 is selected in the same way as the \ttbar signal region but the leptons are required to have 
 same-sign electric charges.

An additional uncertainty is evaluated for the $Wt$ process by replacing the nominal DR sample
with a DS sample, as discussed in Section~\ref{sec:data_mc}, and taking the difference between
the two as the systematic uncertainty.

\subsection{Detector modelling uncertainties}

Systematic uncertainties due to the modelling of the detector response
affect the signal reconstruction efficiency, the unfolding
procedure, and the background estimation. In order to evaluate their
impact, the full analysis is repeated with variations of the detector
modelling and the difference between the nominal and the shifted
results is taken as the systematic uncertainty. 

The uncertainties due to lepton isolation, trigger, identification,
and reconstruction requirements are evaluated
in 2015 data using a tag-and-probe method in leptonically decaying
$Z$-boson events~\cite{Aad:2016jkr}. These
uncertainties are summarised as ``Lepton'' in
Tables~\ref{tab:syst_small_top_pt} and \ref{tab:syst_small_top_pt_normalised}.

The uncertainties due to the jet energy scale and resolution are extrapolated
to $\sqrt{s} = 13$~TeV using a combination of test beam data,
simulation and $\sqrt{s} = 8$~TeV dijet data~\cite{ATL-PHYS-PUB-2015-015}. 
To account for potential mismodelling of the JVT distribution in simulation,
a 2\% systematic uncertainty is applied to the jet efficiency. These
uncertainties are summarised as ``Jet'' in
Tables~\ref{tab:syst_small_top_pt} and
\ref{tab:syst_small_top_pt_normalised}.
Uncertainties due to $b$-tagging, summarised under ``$b$-tagging'',
 are determined using $\sqrt{s}=8$~TeV data as described in Ref.~\cite{ATLAS-CONF-2014-004} for $b$-jets and Ref.~\cite{ATLAS-CONF-2014-046} for $c$- and light-jets, with additional uncertainties to account for the presence of the new Insertable B-Layer detector and the extrapolation from $\sqrt{s}=8$~TeV to $\sqrt{s}=13$~TeV~\cite{ATL-PHYS-PUB-2015-022}.

The systematic uncertainty due to the track-based terms (i.e. those tracks
not associated with other reconstructed objects such as leptons and jets) used in the
calculation of \met\ is evaluated by 
comparing the \met in $Z\rightarrow\mu\mu$ events, which do not contain prompt neutrinos from the hard process, using different generators.
Uncertainties associated with energy scales and resolutions of leptons
and jets are propagated to the \met calculation.

The uncertainty due to the integrated luminosity is $\pm 2.1$\%. 
It is derived, following a methodology similar to that detailed in Ref.~\cite{Aad:1517411}, from a calibration of the luminosity scale using $x$--$y$ beam-separation scans performed in August 2015.
The uncertainty in the pile-up reweighting is evaluated by varying the scale factors by
$\pm1\sigma$ based on the reweighting of the average number of interactions per bunch crossing. 

The uncertainties due to lepton and \met\ modelling are not large for any observable.
For the absolute cross-sections, the uncertainty due to luminosity is not a dominant
systematic uncertainty,  and this uncertainty mainly
cancels in the normalised cross-sections. The luminosity uncertainty does not
 cancel fully since it affects the background subtraction. The uncertainty due to jet
energy scale and JVT is a significant source of uncertainty in the absolute
cross-sections and in some of the normalised cross-sections such as for
$p_{\textrm T}(t\bar{t})$. The uncertainties due to the limited number of MC events
 are evaluated using pseudo-experiments. The data
statistical uncertainty is evaluated using the full covariance matrix
from the unfolding. 

\input{tables/systematics_top_pt.tex}
\input{tables/systematics_top_pt_normalised.tex}

%% file: tables/systematics_top_pt.tex
\begin{table}
\scriptsize
\centering
\begin{tabular}{l  c  c  c  c  c }
\toprule 
$p_{\textrm T}(t)$ & 0 -- 70 GeV & 70 -- 150 GeV & 150 -- 250 GeV & 250 -- 400 GeV & 400 -- 1000 GeV\\
\midrule 
Source & \multicolumn{5}{c}{ Systematic uncertainty (\%) } \\
\midrule 
Radiation scale & $+4.0$  $-3.9$  & $+1.1$  $-3.9$  & $+1.9$  $-3.5$  & $+1.4$  $-5.0$  & $+5.0$  $-5.4$ \\ 
MC generator & $\mp0.9$ & $\mp1.2$ & $\mp1.4$ & $\pm1.6$ & $\mp6.7$\\ 
PDF extrapolation & $\mp2.9$ & $\mp2.8$ & $\mp1.9$ & $\mp0.3$ & $\mp2.4$\\ 
PDF4LHC 100 & $\pm2.2$ & $\pm2.5$ & $\pm2.8$ & $\pm3.7$ & $\pm6.1$\\ 
Parton shower & $\mp8.0$ & $\mp7.7$ & $\mp3.9$ & $\pm3.1$ & $\pm34$\\ 
Background & $+0.3$  $-0.5$  & $+0.2$  $-0.4$  & $\pm 0.2$  & $\pm 0.2$  & $+0.4$  $-1.5$ \\ 
Pile-up & $+0.7$  $-1.4$  & $+0.2$  $-0.6$  & $+0.0$  $-0.4$  & $+ 0.0 $  $-0.4$  & $+4.1$ $-0.0 $\\ 
Lepton & $+0.8$  $-0.7$  & $\pm 0.8$  & $\pm 1.0$  & $\pm 1.6$  & $+3.2$  $-3.0$ \\ 
$b$-tagging & $+3.1$  $-3.6$  & $+3.4$  $-3.9$  & $+3.4$  $-4.0$  & $+4.0$  $-4.7$  & $+6.2$  $-7.2$ \\ 
Jet & $\pm 2.8$  & $+2.6$  $-3.4$  & $+2.0$  $-1.8$  & $+1.9$  $-1.1$  & $+4.5$  $-5.1$ \\ 
$E^{\textrm miss}_{\textrm T}$ & $+0.2$  $-0.1$  & $\pm 0.1$  & $+0.2$  $-0.1$  & $+0.3$  $-0.5$  & $+1.0$  $-0.3$ \\ 
Luminosity & $+2.0$  $-2.1$  & $+2.1$  $-2.2$  & $+2.1$  $-2.2$  & $+2.3$  $-2.4$  & $+3.0$  $-3.1$ \\ 
MC stat. unc. & $\pm0.4$ & $\pm0.3$ & $\pm0.5$ & $\pm0.9$ & $\pm3.2$\\ 
\midrule 
Total syst. unc. & $+11$ $-11$ & $+9$ $-11$ & $+7.3$ $-8.1$ & $+7.5$ $-9.1$ & $+37$ $-37$\\ 
\midrule 
Data statistics & $\pm1.8$ & $\pm1.3$ & $\pm1.8$ & $\pm3.4$ & $\pm10$\\ 
\midrule 
Total uncertainty & $+11$ $-11$ & $+10$ $-11$ & $+7.5$ $-8.3$ & $+8.2$ $-9.8$ & $+38$ $-39$\\ 
\bottomrule 
\end{tabular}
\caption{Summary of the sources of uncertainty in the absolute fiducial differential cross-section as a function of $p_{\textrm T}(t)$. The uncertainties are presented as a percentage of the measured cross-section in each bin. Entries with 0.0 are uncertainties that are less than 0.05 in magnitude. For systematic uncertainties that have only one variation, $\pm$($\mp$) indicate that the systematic shift is positive (negative) and then symmetrised. All uncertainties are rounded to two digits.}
\label{tab:syst_small_top_pt}
\end{table}

%% file: tables/systematics_top_pt_normalised.tex
\begin{table}
\scriptsize
\centering
\begin{tabular}{l  c  c  c  c  c }
\toprule 
$p_{\textrm T}(t)$ & 0 -- 70 GeV & 70 -- 150 GeV & 150 -- 250 GeV & 250 -- 400 GeV & 400 -- 1000 GeV\\
\midrule 
Source & \multicolumn{5}{c}{ Systematic uncertainty (\%) }\\ 
\midrule 
Radiation scale & $+2.1$  $-0.3$  & $+0.0$  $-1.1$  & $+0.4$  $-0.3$  & $+ 0.0 $  $-1.2$  & $+2.1$ $-0.0 $\\ 
MC generator & $\pm0.2$ & $\mp0.2$ & $\mp0.4$ & $\pm2.7$ & $\mp5.4$\\ 
PDF extrapolation & $\mp0.5$ & $\mp0.4$ & $\pm0.4$ & $\pm2.4$ & $\pm0.8$\\ 
PDF4LHC 100 & $\pm0.6$ & $\pm0.3$ & $\pm0.5$ & $\pm1.7$ & $\pm4.0$\\ 
Parton shower & $\mp2.8$ & $\mp2.1$ & $\pm1.6$ & $\pm8.9$ & $\pm41$\\ 
Background & $+0.1$  $-0.2$  & $+0.0$  $-0.1$  & $+0.3$  $-0.0$  & $+0.3$  $-0.1$  & $+0.1$  $-1.2$ \\ 
Pile-up & $+0.4$  $-0.8$  & $\pm 0.0$  & $+0.3$  $-0.2$  & $+0.8$  $-0.7$  & $+5.1$ $-0.0 $\\ 
Lepton & $+0.4$  $-0.3$  & $+0.1$  $-0.3$  & $+0.3$  $-0.1$  & $\pm 0.7$  & $+2.3$  $-1.9$ \\ 
$b$-tagging & $\pm 0.2$  & $\pm 0.2$  & $\pm 0.2$  & $\pm 0.9$  & $+2.3$  $-2.4$ \\ 
Jet & $+0.9$  $-0.8$  & $+0.4$  $-1.0$  & $+0.8$  $-0.6$  & $+3.0$  $-2.4$  & $+6.9$  $-7.3$ \\ 
$E^{\textrm miss}_{\textrm T}$ & $+0.2$  $-0.1$  & $+0.0$  $-0.1$  & $+0.2$  $-0.1$  & $+0.3$  $-0.5$  & $+1.0$  $-0.4$ \\ 
Luminosity & $\pm 0.0$  & $\pm 0.0$  & $\pm 0.0$  & $\pm 0.0$  & $\pm 0.0$ \\ 
MC stat. unc. & $\pm0.0$ & $\pm0.2$ & $\pm0.0$ & $\pm0.4$ & $\pm2.6$\\ 
\midrule 
Total syst. unc. & $+3.8$ $-3.2$ & $+2.2$ $-2.7$ & $+2.1$ $-2.0$ & $+10$ $-10$ & $+42$ $-42$\\ 
\midrule 
Data statistics & $\pm1.8$ & $\pm1.3$ & $\pm1.8$ & $\pm3.4$ & $\pm10$\\ 
\midrule 
Total uncertainty & $+4.2$ $-3.6$ & $+2.6$ $-2.9$ & $+2.8$ $-2.7$ & $+11$ $-11$ & $+44$ $-43$\\ 
\bottomrule 
\end{tabular}
\caption{Summary of the sources of uncertainty in the normalised fiducial differential cross-section as a function of $p_{\textrm T}(t)$. The uncertainties are presented as a percentage of the measured cross-section in each bin. Entries with 0.0 are uncertainties that are less than 0.05 in magnitude. For systematic uncertainties that have only one variation, $\pm$($\mp$) indicate that the systematic shift is positive (negative) and then symmetrised. All uncertainties are rounded to two digits.}
\label{tab:syst_small_top_pt_normalised}
\end{table}

%% file: sections/results.tex
\section{Results}
\label{sec:results}

The unfolded particle-level distributions for the absolute and
normalised fiducial differential cross-sections are presented in
Table~\ref{tab:results}. The total systematic uncertainties
include all sources discussed in Section~\ref{sec:systematics}.

The unfolded normalised data are used to compare with different
generator predictions. The significance of the differences of various
generators, with respect to the data in each observable, are evaluated by calculating
the $\chi^2$ and determining $p$-values using the number of degrees of
freedom (NDF). The $\chi^2$ is determined using:
\begin{equation}
\chi^2 = S^{\textrm T}_{(N - 1)} \cdot \text{Cov}^{-1}_{(N - 1)} \cdot S_{(N - 1)} \text{,}
\end{equation}
where $\text{Cov}^{-1}$ is the inverse of the full bin-to-bin covariance matrix,
including all statistical and systematic uncertainties, $N$ is the
number of bins, and $S$ is a column vector of the differences between the unfolded data and the
prediction. The NDF is equal to the number of bins minus one in the
observable for the normalised cross-sections. In $\text{Cov}$ and
$S$, a single bin is removed from the calculation to account for the
normalisation of the observable, signified by the $(N - 1)$ subscript. The $\chi^2$, NDF, and associated 
$p$-values are presented in Table~\ref{tab:chi2_normalised} for the
normalised cross-sections. Most generators studied agree with the
unfolded data in each observable within the experimental
uncertainties, with the exception of the \POWHEG-Box + \HERWIGpp MC simulation, which
differs significantly from the data in both $p_{\textrm T}(t)$ and
$m(t\bar{t})$.

\begin{figure}
\centering
\includegraphics[width=0.48\textwidth]{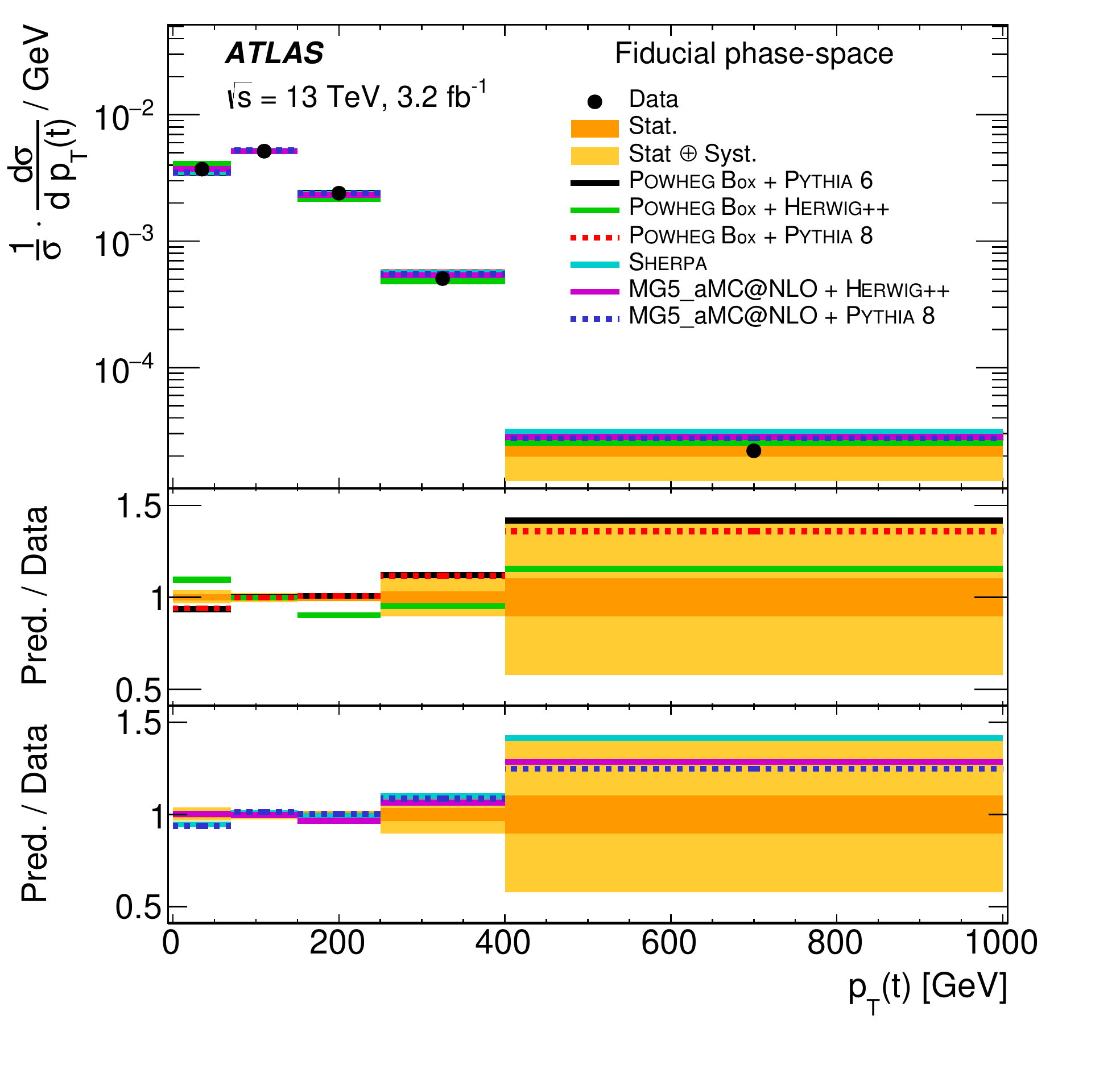}
\includegraphics[width=0.48\textwidth]{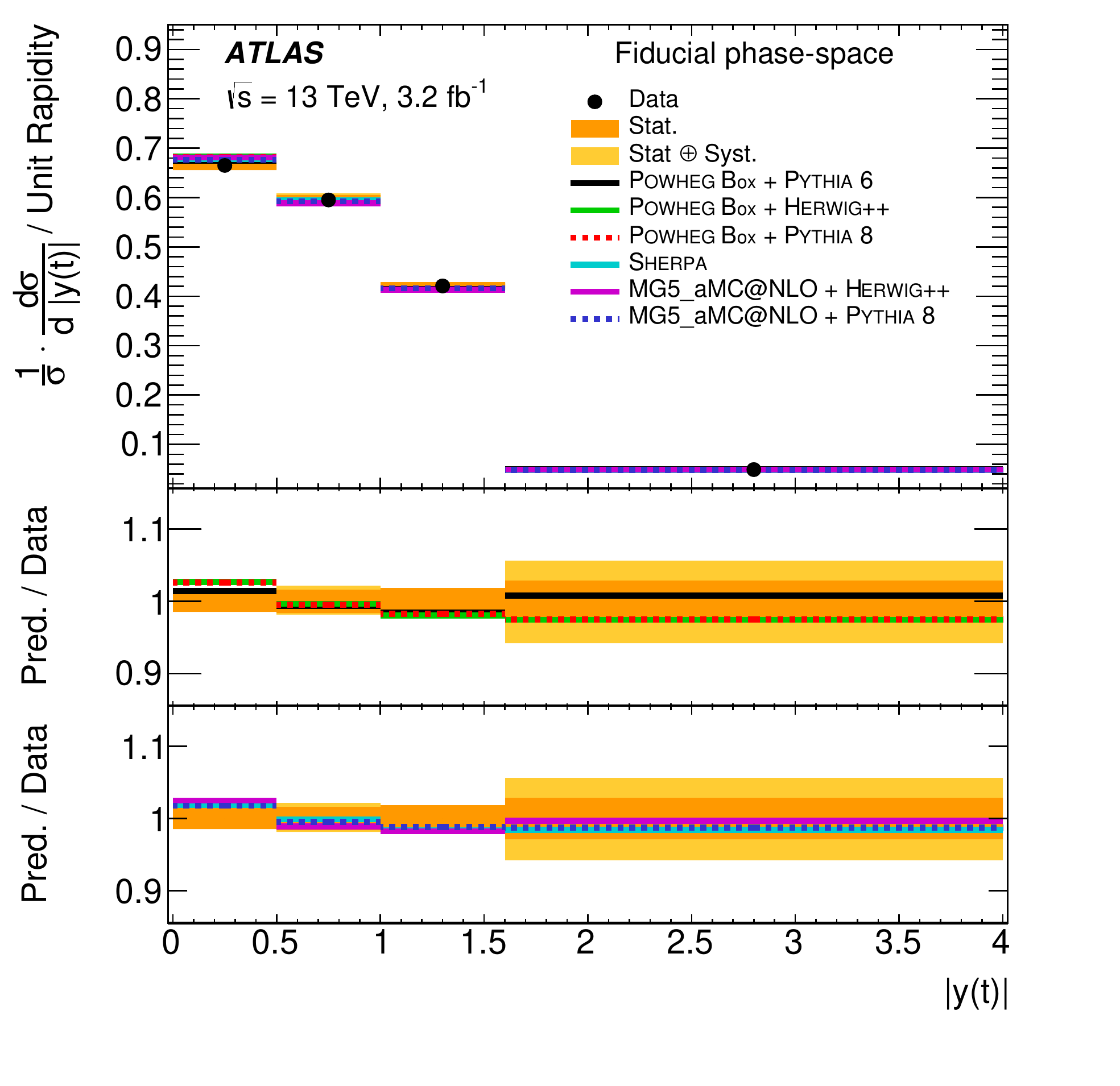}
\includegraphics[width=0.48\textwidth]{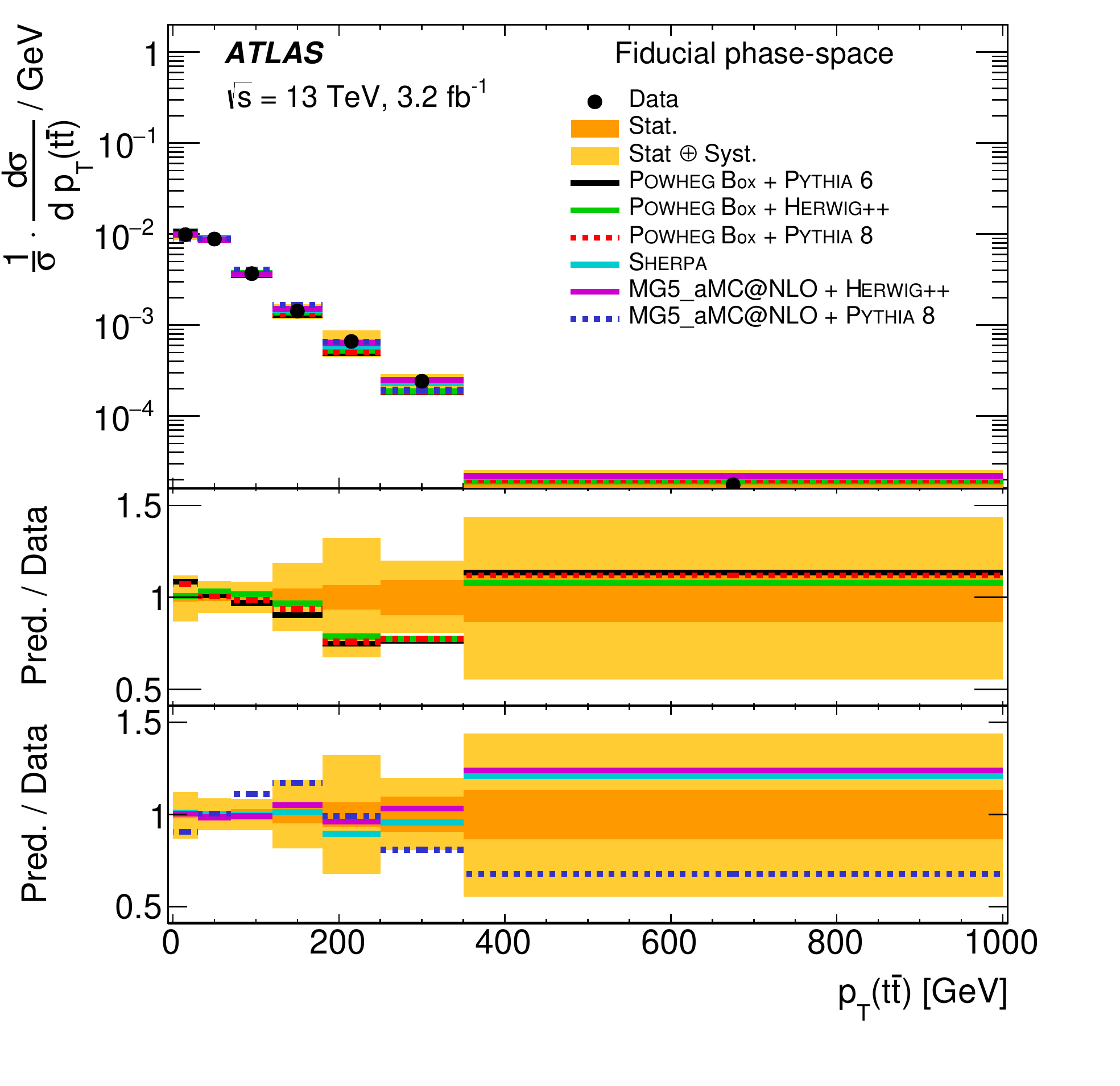}
\includegraphics[width=0.48\textwidth]{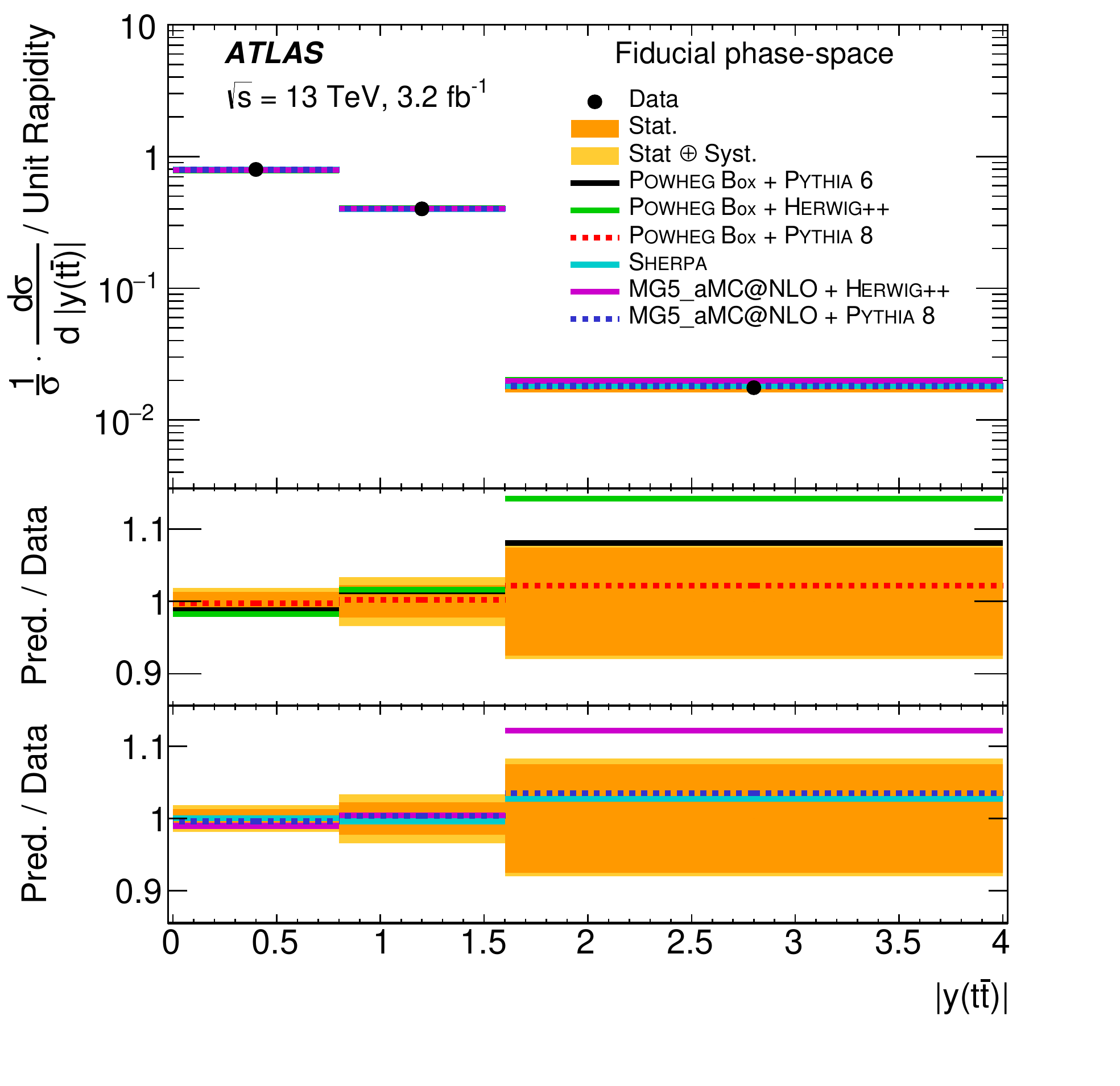}
\includegraphics[width=0.48\textwidth]{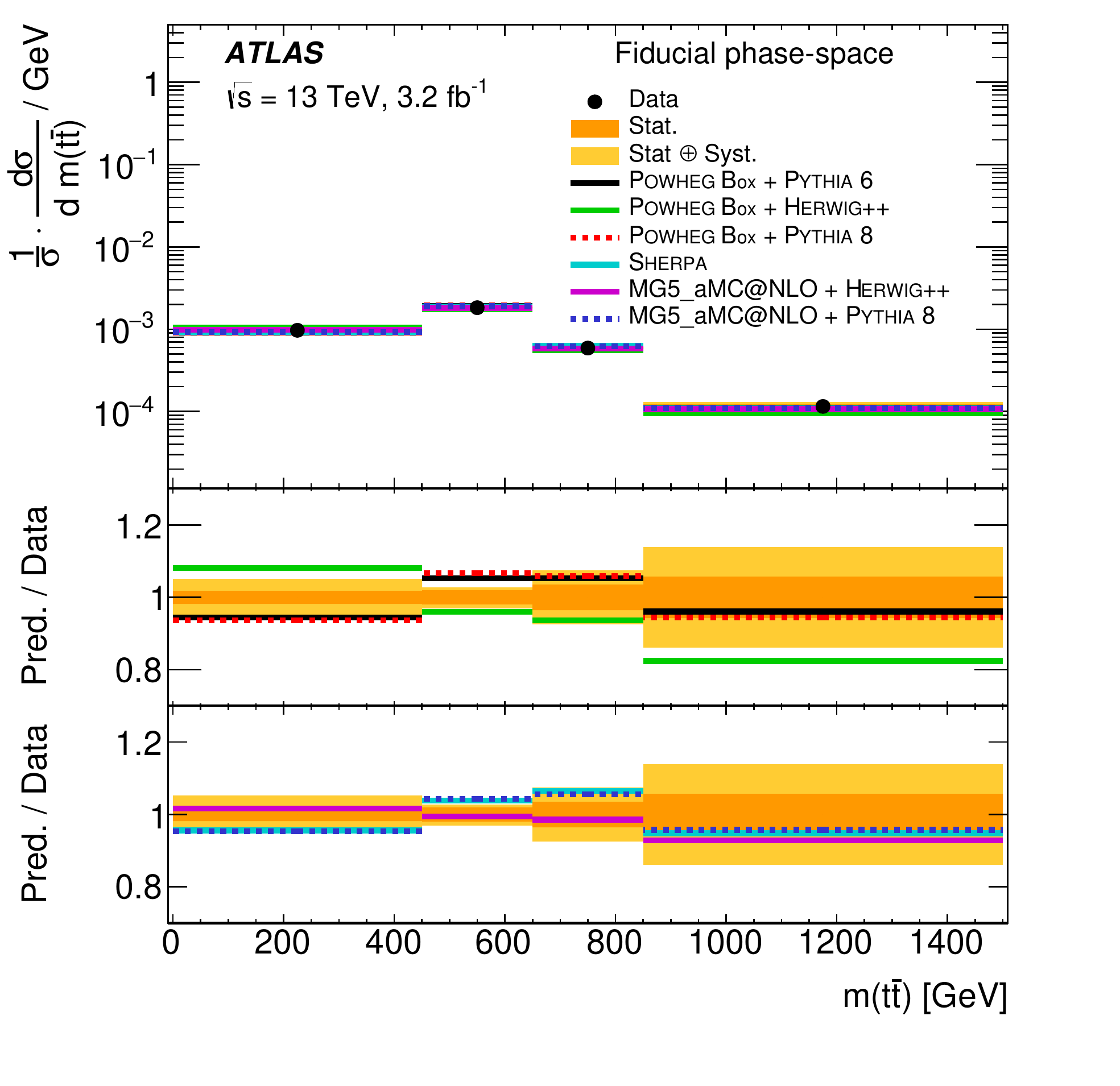}
\caption{The measured normalised fiducial differential cross-sections compared to
  predictions from \POWHEG-Box (\emph{top ratio panel}), \MGatNLO, and \SHERPA (\emph{bottom ratio panel}) interfaced to
  various parton shower programs.}
\label{fig:results}
\end{figure}

The normalised differential cross-sections for all observables are compared
to predictions of different MC generators in Figure~\ref{fig:results}. 

The \POWHEG-Box generator tends to predict a harder $p_{\textrm T}(t)$ spectrum
for the top quark than is observed in data, although the data are still
consistent with the prediction within the experimental
uncertainties. The \MGatNLO~generator appears to agree better with the
observed  $p_{\textrm T}(t)$ spectrum, particularly when interfaced to 
\HERWIGpp. For the $p_{\textrm T}(t\bar{t})$ spectrum, again little
difference is observed between \POWHEG-Box + \mbox{\PYTHIA\!6} and \mbox{\PYTHIA\!8}, and
both generally predict a softer spectrum than the data but are also
consistent within the experimental uncertainties. The \MGatNLO~generator, interfaced to \mbox{\PYTHIA\!8} or \HERWIGpp seems to agree with the data at low to medium values of
$p_{\textrm T}$ but \MGatNLO~+ \HERWIGpp disagrees at higher values. 
For the $m(t\bar{t})$ observable, although the uncertainties are quite large,
 predictions from \POWHEG-Box interfaced to \mbox{\PYTHIA\!6} or \mbox{\PYTHIA\!8} 
 and the \MGatNLO~+ \mbox{\PYTHIA\!8} prediction seem higher than the observed data around 600 \GeV. 
For the rapidity observables, all MC predictions appear to agree
with the observed data, except for the high $|y(t\bar{t})|$ region, where
some of the predictions are slightly higher than the data.

\input{tables/results_summary_all.tex}

\begin{landscape}
\input{tables/chi2_values_normalised.tex}
\end{landscape}

%% file: tables/results_summary_all.tex
\begin{table}
\centering
\begin{tabular}{c c c c c c c}
\toprule
$X$  & $\frac{\text{d}\sigma_{t\bar{t}}}{\text{d}X}$~$[\frac{\text{pb}}{\text{GeV}}]$& $\frac{1}{\sigma_{t\bar{t}}}\frac{\text{d}\sigma_{t\bar{t}}}{\text{d}X}$~$[\frac{1}{\text{GeV}}]$ & Stat. (abs.) & Stat. (norm.) &  Syst. (abs.) & Syst. (norm.) \\
\bottomrule
$p_{\textrm T}(t)$ [GeV] &       &      &   [\%] & [\%] & [\%] & [\%]         \T\B \\
\midrule
0   -- 70   & 7.1   &  0.371  & $\pm$ 1.8 & $\pm$ 1.7 & +11 -11 &  +4 -3.2 \\
70  -- 150  & 9.9   &  0.515  & $\pm$ 1.3 & $\pm$ 1.2 & +10 -11 &  +2.3 -2.7 \\
150 -- 250  & 4.61  &  0.239  & $\pm$ 1.8 & $\pm$ 1.7 &  +7  -8 &  +2.1 -2.0 \\
250 -- 400  & 0.97  &  0.051  & $\pm$ 3.4 & $\pm$ 3.3 &  +7  -9 &  +10  -11 \\
400 -- 1000 & 0.042 &  0.0022 & $\pm$ 10  & $\pm$ 9   & +40 -40 &  +40  -40 \\
\midrule
$p_{\textrm T}(t\bar{t})$ [GeV] &       &       &      &               &               \\
\midrule
0   -- 30   & 9.6    &  0.99   &  $\pm$ 2.2 & $\pm$ 2.0 & +15 -16 & +12 -13 \\
30  -- 70   & 8.6    &  0.88   &  $\pm$ 1.9 & $\pm$ 1.7 &  +8  -8 &  +9 -9  \\
70  -- 120  & 3.6    &  0.368  &  $\pm$ 3.0 & $\pm$ 2.7 & +10 -11 &  +8 -9  \\
120 -- 180  & 0.139  &  0.143  &  $\pm$ 5   & $\pm$ 5   & +24 -24 & +19 -18 \\
180 -- 250  & 0.064  &  0.066  &  $\pm$ 7   & $\pm$ 6   & +40 -40 & +32 -32 \\
250 -- 350  & 0.023  &  0.024  &  $\pm$ 10  & $\pm$ 9   & +24 -24 & +30 -19 \\
350 -- 1000 & 0.0017 &  0.0018 &  $\pm$ 14  & $\pm$ 13  & +50 -50 & +40 -40 \\
\midrule
$m(t\bar{t})$ [GeV] &       &       &      &               &      \\
\midrule
0    -- 450 &  0.94   &  0.097   & $\pm$ 1.8 & $\pm$ 1.6 &  +12 -13 &  +5 -5 \\
450  -- 650 &  1.76   &  0.183   & $\pm$ 2.0 & $\pm$ 1.9 &  +8 -9   &  +2.8 -3.0 \\
650  -- 850 &  0.57   &  0.059   & $\pm$ 4   & $\pm$ 3.3 & +10 -12  & +8 -8 \\
850  -- 1500&  0.111  &  0.0115  & $\pm$ 6   & $\pm$ 5   & +11 -11  & +14 -14 \\
\toprule
$X$  & $\frac{\text{d}\sigma_{t\bar{t}}}{\text{d}X}$~[pb] & $\frac{1}{\sigma_{t\bar{t}}}\frac{\text{d}\sigma_{t\bar{t}}}{\text{d}X}$& Stat. (abs.) & Stat. (norm.) &  Syst. (abs.) & Syst. (norm.) \\
\bottomrule
$|y(t\bar{t}$)|&       &       &      [\%] & [\%] & [\%] & [\%]         \T\B \\ 
\midrule
0.0 -- 0.8  &  7.7   &  0.797  &  $\pm$ 1.3 & $\pm$ 1.1 &  +8  -9  &  +1.8  -1.8 \\
0.8 -- 1.6  &  3.9   &  0.400  &  $\pm$ 2.2 & $\pm$ 2.0 &  +9  -10 &  +3.4  -3.4 \\
1.6 -- 4.0  &  0.170 &  0.0176 &  $\pm$ 7   & $\pm$ 7   & +13 -13 & +8 -8 \\
\toprule
$|y(t)|$   &       &       &      &               &               \\
\midrule
0.0 -- 0.5  & 12.9  &  0.665  & $\pm$ 1.5 & $\pm$ 1.4 &  +8 -10 &  +1.0 -1.3 \\
0.5 -- 1.0  & 11.5  &  0.595  & $\pm$ 1.6 & $\pm$ 1.5 & +10 -10 &  +2.2 -1.9 \\
1.0 -- 1.6  &  8.1  &  0.421  & $\pm$ 1.8 & $\pm$ 1.7 &  +8  -9 &  +1.4 -1.2 \\
1.6 -- 4.0  &  0.95 &  0.0489 & $\pm$ 2.9 & $\pm$ 2.7 &  +8  -9 &  +6  -6 \\

\bottomrule
\end{tabular}
\caption{Summary of the measured absolute
  ($\frac{\text{d}\sigma_{t\bar{t}}}{\text{d}X}$) and normalised
  ($\frac{1}{\sigma_{t\bar{t}}}\frac{\text{d}\sigma_{t\bar{t}}}{\text{d}X}$) differential
  cross-sections, along with the relative statistical (Stat.) and
  systematic (Syst.)
  uncertainties for both the absolute (abs.) and normalised (norm.)
  cross-sections. The results and uncertainties are rounded according
  to recommendations from the Particle Data Group (PDG).}
\label{tab:results}
\end{table}

%% file: tables/chi2_values_normalised.tex
\begin{table}
\scriptsize
\centering
\begin{tabular}{l | c c  | c c  | c c  | c c  | c c }
\toprule
  & \multicolumn{2}{c}{$p_{\text{T}}(t)$}& \multicolumn{2}{c}{$|y(t)|$}& \multicolumn{2}{c}{$p_{\text{T}}(t\bar{t})$}& \multicolumn{2}{c}{$|y(t\bar{t})|$}& \multicolumn{2}{c}{$m(t\bar{t})$}\\
Predictions & $\chi^2$/NDF & $p$-value & $\chi^2$/NDF & $p$-value & $\chi^2$/NDF & $p$-value & $\chi^2$/NDF & $p$-value & $\chi^2$/NDF & $p$-value \\
\midrule
\textsc{Powheg} + \textsc{Pythia} 6          & 5.2/4  & 0.27 & 0.5/3 & 0.92 & 5.5/6 & 0.48 & 0.6/2 & 0.74 & 3.9/4 & 0.42\\
\textsc{Powheg} + \textsc{Pythia} 8          & 4.6/4  & 0.33 & 1.3/3 & 0.73 & 5.1/6 & 0.53 & 0.0/2 & 1.00 & 5.7/4 & 0.22\\
\textsc{Powheg} + \textsc{Herwig++}          & 14.6/4 & 0.01 & 1.4/3 & 0.71 & 4.1/6 & 0.66 & 1.0/2 & 0.61 & 12.0/4& 0.02\\
\textsc{MG5}\_aMC@NLO + \textsc{Herwig++}    & 2.0/4  & 0.74 & 1.3/3 & 0.73 & 0.6/6 & 1.00 & 0.2/2 & 0.90 & 0.9/4 & 0.92\\
\textsc{MG5}\_aMC@NLO + \textsc{Pythia} 8    & 3.6/4  & 0.46 & 0.6/3 & 0.90 & 10.7/6& 0.10 & 0.1/2 & 0.95 & 2.7/4 & 0.61\\
\textsc{Sherpa}                              & 3.8/4  & 0.43 & 0.8/3 & 0.85 & 0.7/6 & 0.99 & 0.0/2 & 1.00 & 2.3/4 & 0.68\\
\textsc{Powheg} + \textsc{Pythia} 6 (radHi)  & 7.8/4  & 0.10 & 0.6/3 & 0.90 & 0.9/6 & 0.99 & 0.4/2 & 0.82 & 3.8/4 & 0.43\\
\textsc{Powheg} + \textsc{Pythia} 6 (radLow) & 5.5/4  & 0.24 & 0.8/3 & 0.85 & 9.6/6 & 0.14 & 0.8/2 & 0.67 & 4.5/4 & 0.34\\
\bottomrule
\end{tabular}
\caption{$\chi^2$ values between the normalised unfolded fiducial cross-section and various predictions from the MC simulation. The number of degrees of freedom (NDF) is equal to one less than the number of bins in the distribution. \textsc{Powheg} refers to \textsc{Powheg}-Box v2.}
\label{tab:chi2_normalised}
\end{table}

%% file: sections/conclusions.tex
\section{Conclusions}
\label{sec:conclusions}

Absolute and normalised differential top-quark pair-production
cross-sections in a fiducial phase-space region are measured using 3.2~fb$^{-1}$ of $\sqrt{s} = 13$~\TeV\ proton--proton 
collisions recorded by the ATLAS detector at the LHC in 2015. The differential 
cross-sections are determined in the $e^{\pm}\mu^{\mp}$ channel, for the
transverse momentum and the absolute
rapidity of the top quark, as well as the transverse momentum, the
absolute rapidity, and the invariant mass of the top-quark pair. 
The measured differential cross-sections are compared
to predictions of NLO generators matched to parton showers and the results are
found to be consistent with all models within the experimental
uncertainties, with the exception of \POWHEG\!-Box + \Herwig\!++, which
deviates from the data in the $p_{\textrm T}(t)$ and $m(t\bar{t})$ observables.

%% file: acknowledgements/Acknowledgements.tex

We thank CERN for the very successful operation of the LHC, as well as the
support staff from our institutions without whom ATLAS could not be
operated efficiently.

We acknowledge the support of ANPCyT, Argentina; YerPhI, Armenia; ARC, Australia; BMWFW and FWF, Austria; ANAS, Azerbaijan; SSTC, Belarus; CNPq and FAPESP, Brazil; NSERC, NRC and CFI, Canada; CERN; CONICYT, Chile; CAS, MOST and NSFC, China; COLCIENCIAS, Colombia; MSMT CR, MPO CR and VSC CR, Czech Republic; DNRF and DNSRC, Denmark; IN2P3-CNRS, CEA-DSM/IRFU, France; SRNSF, Georgia; BMBF, HGF, and MPG, Germany; GSRT, Greece; RGC, Hong Kong SAR, China; ISF, I-CORE and Benoziyo Center, Israel; INFN, Italy; MEXT and JSPS, Japan; CNRST, Morocco; NWO, Netherlands; RCN, Norway; MNiSW and NCN, Poland; FCT, Portugal; MNE/IFA, Romania; MES of Russia and NRC KI, Russian Federation; JINR; MESTD, Serbia; MSSR, Slovakia; ARRS and MIZ\v{S}, Slovenia; DST/NRF, South Africa; MINECO, Spain; SRC and Wallenberg Foundation, Sweden; SERI, SNSF and Cantons of Bern and Geneva, Switzerland; MOST, Taiwan; TAEK, Turkey; STFC, United Kingdom; DOE and NSF, United States of America. In addition, individual groups and members have received support from BCKDF, the Canada Council, CANARIE, CRC, Compute Canada, FQRNT, and the Ontario Innovation Trust, Canada; EPLANET, ERC, ERDF, FP7, Horizon 2020 and Marie Sk{\l}odowska-Curie Actions, European Union; Investissements d'Avenir Labex and Idex, ANR, R{\'e}gion Auvergne and Fondation Partager le Savoir, France; DFG and AvH Foundation, Germany; Herakleitos, Thales and Aristeia programmes co-financed by EU-ESF and the Greek NSRF; BSF, GIF and Minerva, Israel; BRF, Norway; CERCA Programme Generalitat de Catalunya, Generalitat Valenciana, Spain; the Royal Society and Leverhulme Trust, United Kingdom.

The crucial computing support from all WLCG partners is acknowledged gratefully, in particular from CERN, the ATLAS Tier-1 facilities at TRIUMF (Canada), NDGF (Denmark, Norway, Sweden), CC-IN2P3 (France), KIT/GridKA (Germany), INFN-CNAF (Italy), NL-T1 (Netherlands), PIC (Spain), ASGC (Taiwan), RAL (UK) and BNL (USA), the Tier-2 facilities worldwide and large non-WLCG resource providers. Major contributors of computing resources are listed in Ref.~\cite{ATL-GEN-PUB-2016-002}.

%% file: atlas_authlist.tex
\begin{flushleft}
{\Large The ATLAS Collaboration}

\bigskip

M.~Aaboud$^\textrm{\scriptsize 136d}$,
G.~Aad$^\textrm{\scriptsize 87}$,
B.~Abbott$^\textrm{\scriptsize 114}$,
J.~Abdallah$^\textrm{\scriptsize 8}$,
O.~Abdinov$^\textrm{\scriptsize 12}$,
B.~Abeloos$^\textrm{\scriptsize 118}$,
R.~Aben$^\textrm{\scriptsize 108}$,
O.S.~AbouZeid$^\textrm{\scriptsize 138}$,
N.L.~Abraham$^\textrm{\scriptsize 152}$,
H.~Abramowicz$^\textrm{\scriptsize 156}$
% ...
\bigskip
\\
% ...
$^{8}$ Department of Physics, The University of Texas at Arlington, Arlington TX, United States of America\\
% ...
$^{12}$ Institute of Physics, Azerbaijan Academy of Sciences, Baku, Azerbaijan\\
% ...
$^{87}$ CPPM, Aix-Marseille Universit{\'e} and CNRS/IN2P3, Marseille, France\\
% ...
$^{108}$ Nikhef National Institute for Subatomic Physics and University of Amsterdam, Amsterdam, Netherlands\\
% ...
$^{114}$ Homer L. Dodge Department of Physics and Astronomy, University of Oklahoma, Norman OK, United States of America\\
% ...
$^{118}$ LAL, Univ. Paris-Sud, CNRS/IN2P3, Universit{\'e} Paris-Saclay, Orsay, France\\
% ...
$^{136}$ $^{(a)}$ Facult{\'e} des Sciences Ain Chock, R{\'e}seau Universitaire de Physique des Hautes Energies - Universit{\'e} Hassan II, Casablanca; $^{(b)}$ Centre National de l'Energie des Sciences Techniques Nucleaires, Rabat; $^{(c)}$ Facult{\'e} des Sciences Semlalia, Universit{\'e} Cadi Ayyad, LPHEA-Marrakech; $^{(d)}$ Facult{\'e} des Sciences, Universit{\'e} Mohamed Premier and LPTPM, Oujda; $^{(e)}$ Facult{\'e} des sciences, Universit{\'e} Mohammed V, Rabat, Morocco\\
% ...
$^{138}$ Santa Cruz Institute for Particle Physics, University of California Santa Cruz, Santa Cruz CA, United States of America\\
% ...
$^{152}$ Department of Physics and Astronomy, University of Sussex, Brighton, United Kingdom\\
% ...
$^{156}$ Raymond and Beverly Sackler School of Physics and Astronomy, Tel Aviv University, Tel Aviv, Israel\\
% ...
$^{a}$ Also at Department of Physics, King's College London, London, United Kingdom\\
$^{b}$ Also at Institute of Physics, Azerbaijan Academy of Sciences, Baku, Azerbaijan\\
$^{c}$ Also at Novosibirsk State University, Novosibirsk, Russia\\
$^{d}$ Also at TRIUMF, Vancouver BC, Canada\\
$^{e}$ Also at Department of Physics {\&} Astronomy, University of Louisville, Louisville, KY, United States of America\\
% ...
\end{flushleft}